\documentclass[aps,twocolumn,showpacs,preprintnumbers,noshowpacs,superscriptaddress,6pt]{revtex4-2}

\usepackage{graphicx}  
\usepackage{subfigure}
\usepackage{multirow}
\usepackage{adjustbox}
\usepackage{natbib}

\usepackage{fancyhdr}
\usepackage{longtable}
\usepackage{parskip}
\usepackage[T1]{fontenc}
\usepackage{dcolumn}   

\usepackage{bm}        
\usepackage{amsfonts}  
\usepackage{amsmath}   
\usepackage{amssymb}   
\usepackage{siunitx}
\usepackage{booktabs}
\usepackage{physics}
\usepackage{mathtools}
\usepackage{flafter}
\usepackage{threeparttable}
\usepackage{subfiles} 
\usepackage{subfigure}
\usepackage{import}
\usepackage{float}

\usepackage{ulem}

\usepackage{colortbl}

\begin{document}

\title{Electron affinity of oganesson}
\author{M.~Y.~Kaygorodov}
\affiliation{Department of Physics, St. Petersburg State University, 7/9 Universitetskaya nab., 199034 St. Petersburg, Russia}

\author{L.~V.~Skripnikov}
\affiliation{B. P. Konstantinov Petersburg Nuclear Physics Institute of National Research Centre “Kurchatov Institute”, Gatchina, 188300 Leningrad District, Russia}
\affiliation{Department of Physics, St. Petersburg State University, 7/9 Universitetskaya nab., 199034 St. Petersburg, Russia}

\author{I.~I.~Tupitsyn}
\affiliation{Department of Physics, St. Petersburg State University, 7/9 Universitetskaya nab., 199034 St. Petersburg, Russia}

\author{E.~Eliav}
\affiliation{School of Chemistry, Tel Aviv University, 69978 Tel Aviv, Israel}

\author{Y.~S.~Kozhedub}
\affiliation{Department of Physics, St. Petersburg State University, 7/9 Universitetskaya nab., 199034 St. Petersburg, Russia}

\author{A.~V.~Malyshev}
\affiliation{Department of Physics, St. Petersburg State University, 7/9 Universitetskaya nab., 199034 St. Petersburg, Russia}

\author{A.~V.~Oleynichenko}
\affiliation{B. P. Konstantinov Petersburg Nuclear Physics Institute of National Research Centre “Kurchatov Institute”, Gatchina, 188300 Leningrad District, Russia}
\affiliation{Department of Chemistry, M. V. Lomonosov Moscow State University, 119991 Moscow, Russia}

\author{V.~M.~Shabaev}
\affiliation{Department of Physics, St. Petersburg State University, 7/9 Universitetskaya nab., 199034 St. Petersburg, Russia}

\author{A.~V.~Titov}
\affiliation{B. P. Konstantinov Petersburg Nuclear Physics Institute of National Research Centre “Kurchatov Institute”,
Gatchina, 188300 Leningrad District, Russia}

\author{A.~V.~Zaitsevskii}
\affiliation{B. P. Konstantinov Petersburg Nuclear Physics Institute of National Research Centre “Kurchatov Institute”,
Gatchina, 188300 Leningrad District, Russia}
\affiliation{Department of Chemistry, M. V. Lomonosov Moscow State University, 119991 Moscow, Russia}

\date{\today}
\begin{abstract}
The electron affinity (EA) of superheavy element Og is calculated by the use of the relativistic Fock-space coupled cluster (FSCC) and configuration interaction methods.
The FSCC cluster operator expansion included single, double, and triple excitations treated in a non-perturbative manner.
The Gaunt and retardation electron-electron interactions are taken into account.
Both methods yield the results that are in agreement with each other.
The quantum electrodynamics correction to EA is evaluated using the model Lamb-shift operator approach.
The electron affinity of Og is obtained to be 0.076(4) eV.
\end{abstract}
\maketitle

\section{Introduction}\label{seq:intro}
Over the last decades, progress in the synthesis of superheavy elements (SHEs) has manifested itself in completing the seventh period of the periodic table \cite{2006_OganessianY_PhysRevC, 2012_OganessianY_PhysRevLett}.
The electronic structure of these elements at the edge of the periodic table poses a challenge for atomic physics.
A large number of electrons coupled with a complex interplay between the relativistic, correlation, and quantum-electrodynamics (QED) effects may result in entirely different physical properties of SHEs compared to their lighter homologous.
These assumptions have been experimentally confirmed recently for the ground state configuration of Lr, which appeared to be different from that of Lu~\cite{2015_SatoT_Nature}.
Hence, theoretical investigations of the properties of SHEs become important for experiments on atomic structure~\cite{2015_SatoT_Nature, 2018_ChhetriP_PhysRevLett, 2018_RaederS_PhysRevLett} and chemical properties~\cite{2002_DullmannC_Nature, 2007_EichlerR_Nature, 2008_EichlerR_AngewChemIntEd, 2010_EichlerR_RadChimActa, 2014_DmitrievS_MendComm, 2017_AksenovN_EurPhysJA} as well as for the concepts of the periodic-table extension~\cite{1971_FrickeB_ThChimAct21, 1996_SeaborgG_JChemSoc, 2006_NefedovV_DoklPhysChem, 2010_PykkoP_PhysChemChemPhys13, 2018_JerabekP_PhysRevLett, 2019_KaygorodovM_POS}.
A number of papers is devoted to calculations of the properties of SHEs by various methods \cite{1970_MannJ_JChemPhys, 2001_LandauA_JChemPhys, 2002_EliavE_JPhysB, Mosyagin:06b, Zaitsevskii:06a, 2007_LiuY_PhysRevA,  2008_DinhT_PhysRevA78, 2008_PershinaV_JChemPhys129, 2009_BorschevskyA_ChemPhysLett, 2009_GoidenkoI_EurPhysJD, 2009_ZaitsevskiiA_RussChemRev, Zaitsevskii:13b, 2014_DemidovY_RussChemBull, 2014_HangeleT_ChemPhysLett, 2016_DzubaV_PhysRevA, 2016_GingesJ_PhysRevA, *2016_GingesJ_JPhysBAtMolOptPhys, 2019_JerabekP_JPhysChemA, 2019_MewesJ_AngewChemIntEd, 2020_LackenbyB_PhysRevA101}; the reader is also referred to the reviews on advances in computational methods for electronic structure of SHEs \cite{2007_IndelicatoP_EurPhysJD, 2015_EliavE_NuclPhysA, 2015_PershinaV_NuclPhysA, 2015_SchwerdtfegerP_NuclPhysA,  2019_PershinaV_RadChimActa, 2020_MosyaginN_IntJQuantChem} and the general reviews on this topic \cite{ 2015_OganessianY_RepProgPhys, 2017_OganessianY_PhysScrip, 2019_GiulianiS_RevModPhys} that address the nuclear aspects of the problem as well.

Great attention within the theoretical studies is paid to the increasing impact of the relativistic effects on the electronic structure of SHEs.
An example of the property that distinguishes SHE from its lighter homologues is established in Oganesson $(\mathrm{Og},\, Z=118)$.
The relativistic calculations within the Dirac-Coulomb-Breit Hamiltonian performed in Ref.~\cite{1996_EliavE_PhysRevLett} have demonstrated that Og has a positive electron affinity (EA), which was evaluated to be $0.056(10)$~eV.
This is in contrast to the nonrelativistic calculations which do not predict any positive EA.
In other words, Og, while having a noble-gas electron configuration, can form a negatively charged ion, which qualitatively differentiates Og from the other noble gases.
In Ref.~\cite{2003_GoidenkoI_PhysRevA}, the QED contribution to EA of Og was calculated to be $-0.0059(5)$~eV; together with an improvement of the electronic correlation result, EA in Og was found to be $0.058(3)$~eV.
In the work devoted to the calculation of the Og excitation spectrum~ \cite{2018_LackenbyB_PhysRevA98_Og}, which also included the QED correction, a value for EA of $0.096$~eV  was obtained.
The $40\%$ discrepancy between these two results has motivated us to reexamine the electronic structure of Og and provide an independent evaluation of EA.
\par
The main obstacle in the calculations of EA for Og is the fact that there is no bound state of Og anion within the framework of the Dirac-Fock (DF) approach.
The formation of the bound state of Og$^-$ occurs through a combination of the relativistic and electron-electron interaction effects beyond the self-consistent field approximation.
In the present work, EA for Og is calculated using two conceptually different methods, namely the Fock-space coupled cluster (FSCC) and configuration interaction (CI) methods.
The effects of inter-electronic interaction are taken into consideration within the Dirac-Coulomb-Breit Hamiltonian, whereas the QED corrections are evaluated exploiting the model QED operator approach.
A special procedure to eliminate the errors associated with the choice of the basis set in the FSCC method is applied.

The paper is organized as follows. In Sec.~\ref{sec:methods} an overview of the methods and main features of their implementation are presented. 
Sec.~\ref{seq:details} is focused on particular aspects of the methods as well as the numerical details.
In Sec.~\ref{seq:discussion} we discuss the results obtained with the FSCC and CI methods and compare them with the previous theoretical predictions.
A summary of the results is given in Sec.~\ref{seq:conclusion}.
Comprehensive details of the FSCC calculations are collected in App.~\ref{seq:appendix_a}.

The atomic units are used throughout the paper.

\section{Methods}\label{sec:methods}
In the present work, the electronic structure of Og is studied by means of two approaches: FSCC and CI.
For the FSCC calculations we mainly use the DIRAC program \cite{2020_SaueT_JChemPhys}.
A part of the correlation effects beyond the level provided by DIRAC is computed using the EXP-T package~\cite{2020_OleynichenkoA_Symmetry}.
The implementation of the CI method in the basis of the Dirac-Fock-Sturm orbitals (CI-DFS) \cite{2003_TupitsynI_OptSpectrosc, 2003_TupitsynI_PhysRevA, 2005_TupitsynI_PhysRevA, 2018_TupitsynI_PhysRevA} is used.
Hereafter, we give a brief description of the methods and introduce the notations.
\par
\subsection{FSCC method}
The FSCC method implies the construction of an effective Hamiltonian in a model space (MS) defined by the choice of ``valence'' one-electron functions.
The MS is decomposed into sectors according to the number of holes/particles relative to a certain closed-shell configuration.
The latter forms a one-dimensional subspace of the MS, which corresponds to the sector $0h0p$.
Single-particle states relative to the sector $0h0p$ form a sector $0h1p$.
The CC correlated wave-function is constructed subsequently for each sector starting from the sector $0h0p$.
EA in the FSCC method is defined as the lowest eigenvalue of the effective Hamiltonian in the $0h1p$ sector.
We adopt the notation $\epsilon$ for EA.

The cluster operator includes single (S), double (D) and eventually triple (T) excitations (FSCC-SD or FSCC-SDT).
The innermost DF spinors are normally kept frozen at the FSCC stage of calculations whereas the highest-energy virtual spinors are fully rejected; we shall use the term ``active space'' (AS) for the linear space spanned by the remainder DF spinors.
The latter are obtained as solutions of the DF equations in a basis of primitive Gaussian functions.
\par
We employ four different Hamiltonians throughout the FSCC calculations: 1) 4-component Dirac-Coulomb Hamiltonian~$H_{\mathrm{DC}}$,
\begin{equation}\label{eq:H^DC}
    H_{\mathrm{DC}} = \Lambda^{+} \Big[ \sum_{i=1}^{N} h_i^{\mathrm{D}} + \sum_{\substack{i=2, \\ j< i}}^N V^{\mathrm{C}}_{ij} \Big] \Lambda^{+}, \qquad V^{\mathrm{C}}_{ij} = \frac{1}{r_{ij}},
\end{equation}
where $h^{\mathrm{D}}$ is the one-electron Dirac Hamiltonian which includes the interaction with the nucleus with the nuclear charge density modeled by a Gaussian distribution, $N$ is the total number of electrons, $r_{ij}$ is the distance between $i$-th and $j$-th electrons, and $\Lambda^{+}$ is the projector on the positive-energy states of the Dirac-Fock Hamiltonian $h^{\mathrm{DF}}$.
2) 2-component Dirac-Coulomb Hamiltonian with the generalized relativistic effective core potential~(GRECP)~$H_{\mathrm{GRECP}}$,
\begin{equation}
    H_{\mathrm{GRECP}} = \sum_{i=N_{\mathrm{c}}+1}^{N} \left( h_i^{\mathrm{S}} + V^{\mathrm{GRECP}}_{i} \right) + \sum_{\substack{i=N_{\mathrm{c}}+1,\\ j< i}}^N  V^{\mathrm{C}}_{ij},
\end{equation}
where $h^{\mathrm{S}}$ is the Schr\"{o}dinger Hamiltonian of a free electron, $ V^{\mathrm{GRECP}}_{i}$ is the GRECP operator from Ref.~\cite{2020_MosyaginN_IntJQuantChem} which models the interaction of a valence or outer-core $i$-th electron with the inner-core electrons and the nucleus, and $N_\mathrm{c}$ is the number of the inner-core electrons; the GRECP operator effectively takes into account the Breit interaction.
3) 2-component Hamiltonian $H_{\mathrm{X2Cmmf}}$~\cite{2009_SikkemaJ_JChemPhys}, the one-electron part of which exactly reproduces the positive-energy spectrum of the parent $h^{\mathrm{DF}}$ Hamiltonian, and two-electron part is represented by the Coulomb operator $V^{\mathrm{C}}$.
4) Hamiltonian $H_{\mathrm{X2Cmmf}}^{\mathrm{G}}$ which differs from $H_{\mathrm{X2Cmmf}}$ by addition of the correction arising from the Gaunt interaction operator $V^{\mathrm{G}}$,
\begin{equation}\label{eq:V^Gaunt}
    \qquad V^{\mathrm{G}}_{ij} = -\frac{(\bm{\alpha}_i \cdot \bm{\alpha}_{j})}{r_{ij}},
\end{equation}
where $\bm{\alpha}$ is a vector incorporating the Dirac matrices, to the one-electron part of the Hamiltonian written in normal order with respect to the Fermi vacuum (i.e. to the Fock operator). 
The difference of the energies obtained with these two Hamiltonians allows us to estimate the effect of the Gaunt interaction at the FSCC level.
For EA this correction reads as
\begin{equation}\label{eq:EA^G}
    \delta \epsilon^{\mathrm{G}} = \epsilon_{\mathrm{X2Cmmf}}^{\mathrm{G}} - \epsilon_{\mathrm{X2Cmmf}},
\end{equation}
where $\epsilon_{\mathrm{X2Cmmf}}^{\mathrm{G}}$ and $\epsilon_{\mathrm{X2Cmmf}}$ are the values of EA calculated with the $H_{\mathrm{X2Cmmf}}$ and $H_{\mathrm{X2Cmmf}}^{\mathrm{G}}$ Hamiltonians, respectively.

\subsection{CI-DFS method}
The essence of the CI method consists in determination of the lowest eigenvalue(s) of the Dirac-Coulomb-Breit Hamiltonian
\begin{equation}\label{eq:H^DCB}
    H_{\mathrm{DCB}} = \Lambda^{+} \Bigg[ \sum_{i=1}^{N} h_i^{\mathrm{D}} + \sum_{\substack{i=2 \\ j< i}}^N \left( V^{\mathrm{C}}_{ij} + V^{\mathrm{G}}_{ij} + V^{\mathrm{R}}_{ij} \right) \Bigg] \Lambda^{+},
\end{equation}
with
\begin{equation}\label{eq:V^Ret}
    \qquad V^{\mathrm{R}}_{ij} = -\frac{1}{2} \left[\frac{(\bm{\alpha}_i \cdot \bm{r}_{ij}) (\bm{\alpha}_j \cdot \bm{r}_{ij})}{r^3_{ij}} - \frac{(\bm{\alpha}_i \cdot \bm{\alpha}_{j})}{r_{ij}}\right]
\end{equation}
being the retardation interaction operator, in the many-electron basis of configuration-state functions (CSFs).
Each CSF is represented as a linear combination of the Slater determinants and is an eigenfunction of the operator $J^2$.
The active space in the CI-DFS method is constructed according to the restricted active-space (RAS) scheme from the eigenfunctions of $h^{\mathrm{DF}}$ (which also determines the $\Lambda^+$ projector in Eq. (\ref{eq:H^DCB})) defined in a combined basis of the DF and Dirac-Fock-Sturm (DFS) orbitals, where $\varphi^{\mathrm{DF}}$ are orbitals for the occupied states and $\varphi^{\mathrm{DFS}}$ are orbitals for the virtual ones, see Refs.~\cite{2003_TupitsynI_OptSpectrosc, 2003_TupitsynI_PhysRevA, 2005_TupitsynI_PhysRevA}.
The orbitals are found as numerical solutions of the DF and DFS equations.
We obtain EA within the CI-DFS method as 
\begin{equation}\label{eq:EA_CI}
    \epsilon_{\mathrm{CI}} = E_{\mathrm{CI}}(\mathrm{Og}) - E_{\mathrm{CI}}(\mathrm{Og^-}),
\end{equation}
where $E_{\mathrm{CI}}(\mathrm{Og})$ and $E_{\mathrm{CI}}(\mathrm{Og^-})$ are the CI energies calculated for Og and Og$^-$ configurations, respectively.
\par
Additionally, the DF and DFS equations can be modified by incorporating a local polarization potential $V^{\mathrm{pol}}$ into the self-consistent procedure.
This modification is motivated by the fact that the DF equations for the $\mathrm{Og}^-$ configuration yield no solution for the $8s$ orbital $\varphi^{\mathrm{DF}}_{8s}$.
To construct $\varphi^{\mathrm{DF}}_{8s}$ at the DF level for the subsequent use in the CI-DFS procedure, we add to $h^{\mathrm{DF}}$, defined for the frozen core of neutral Og, the polarization potential, which represents an attractive interaction of the loosely bound $8s$ electron with the induced dipole moment of the valence shell(s) and partially takes into account the correlation effects at the one-electron level.
There are many possible choices of the polarization potential~$V^{\mathrm{pol}}$ \cite{1973_NorcrossD_PhysRevA, 1970_DalgarnoA_ChemPhysLett, 1977_BaylisW_JPhysBAtomMolPhys}, see also review \cite{2010_MitroyJ_JPhysB}.
In the present work we adopt the most widely used form of the polarization potential proposed in Ref.~\cite{1943_BatesD}
\begin{equation}\label{eq:V^pol}
    V^{\mathrm{pol}}_{8s} = -\frac{\alpha_d}{2(r^2+r_\mathrm{cut}^2)^2},
\end{equation}
where $\alpha_d$ is determined as the dipole polarizability and $r_{\mathrm{cut}}$ is an adjustable cutoff parameter which is related to the average radius of the valence shell.
This potential is included into the DFS equations for the virtual orbitals as well.
Then, $h^{\mathrm{DF}}$ with $V^{\mathrm{pol}}$ introduced is diagonalized in the basis of $\varphi^{\mathrm{DF}}_{8s}$ and virtual orbitals by holding the orthogonality to the core.
We stress that the occupied orbitals in Og are not affected by the inclusion of the polarization potential.
\par
To evaluate the QED correction to EA we use the QED operator approach~\cite{2003_PyykkoP_JPhysB, 2005_FlambaumV_PhysRevA, 2016_GingesJ_PhysRevA, 2013_TupitsynI_OptSpectrosc, 2013_ShabaevV_PhysRevA}.
We incorporate the model QED operator (QEDMOD) $V^{\mathrm{QED}}_{\mathrm{mod}}$ presented in Refs.~\cite{2013_ShabaevV_PhysRevA, 2015_ShabaevV_CompPhysComm, *2018_ShabaevV_CompPhysComm} into the many-electron Hamiltonian $H_{\mathrm{DCB}}$ and perform two series of the calculations: one with $V^{\mathrm{QED}}_{\mathrm{mod}}$ included into the Hamiltonian and the other without it.
The corresponding QED correction to EA is 
\begin{equation}\label{eq:dEA_QED}
    \delta \epsilon^{\mathrm{QED}} = \epsilon_{\mathrm{CI}}^{\mathrm{QED}} - \epsilon_{\mathrm{CI}},
\end{equation}
where $\delta \epsilon^{\mathrm{QED}}_{\mathrm{CI}}$ is EA calculated with the inclusion of the operator $V^{\mathrm{QED}}_{\mathrm{mod}}$ into the CI-DFS equations.
\section{Computational Details}\label{seq:details}
\subsection{FSCC details}\label{subseq:details:fscc}
Standard basis sets optimized for neutral species calculations are inappropriate for describing small electron affinities due to the lack of diffuse functions.
For instance, the FSCC calculations with the Dyall's basis set AAE4Z ($36s36p25d18f12g6h2i$) for Og do not yield positive EA value; therefore this basis set is needed to be customized and augmented for the problem.
\par
Briefly, the results of the optimization are formulated as follows.
In Table~\ref{table:1}, we present the results for EA calculated with the GRECP Hamiltonian and the FSCC-SD method for the basis set subsequently optimized according to the procedure similar to that proposed in Ref.~\cite{Mosyagin:00} and described in Appendix A.
The EA value monotonically increases with basis functions for the higher angular momenta $L$ being optimized.
The most noticeable change in the EA values which is $\Delta \epsilon_{\mathrm{FSCC}}^{\mathrm{SD}}=0.00232(68)$ eV occurs when the $g$-type basis functions are added to the set.
This is due to the fact that they are the most important for describing the polarization effect of the atom as well as the fact that there were no such functions at the previous optimization stages.
The value of EA evaluated with the fully optimized basis set for $H_{\mathrm{GRECP}}$ is 0.0725(14) eV.
\begin{table}[htbp]
\centering

\caption{EA for Og calculated with the FSCC-SD method depending on the number of the optimized angular momenta in the basis.
The notation $\chi^{\lambda}$, where $\lambda=s,\dots,i$, means the basis set which includes the optimized functions with the orbital quantum number $L$ up to $\lambda$.}
\label{table:1}
\begin{tabular}{c S[table-format=1.4(2)]} 

\toprule
\multicolumn{1}{c}{Optimized basis}  &
\multicolumn{1}{c}{$\mathrm{EA}$, eV} \\
\midrule
$\chi^s$ & 0.0678(2) \\
$\chi^p$ & 0.0685(3) \\
$\chi^d$ & 0.0691(5) \\
$\chi^f$ & 0.0698(6) \\
$\chi^g$ & 0.0722(13) \\
$\chi^h$ & 0.0726(14) \\
$\chi^i$ & 0.0725(14) \\
\bottomrule

\end{tabular}
\end{table}
\par
After the basis set has been optimized employing the $H_{\mathrm{GRECP}}$ Hamiltonian, we consider different Hamiltonians to verify the stability of the results.
In Table~\ref{table:2}, the EA values obtained within the FSCC-SD method with the basis set $\chi^i$ for the $H_{\mathrm{GRECP}}$, $H_{\mathrm{X2Cmmf}}$, and $H_{\mathrm{DC}}$ Hamiltonians are presented.
The same number of the correlated electrons and  virtual orbitals as in the optimization procedure is used in all calculations.
The difference between EA calculated with the $H_{\mathrm{GRECP}}$ Hamiltonian and that evaluated with the $H_{\mathrm{DC}}$ Hamiltonian is $0.002$ eV.
The exact 2-component Hamiltonian $H_{\mathrm{X2Cmmf}}$ yields $\mathrm{EA}$ between the results obtained with the $H_{\mathrm{GRECP}}$ and $H_{\mathrm{DC}}$ Hamiltonians.
We note that the deviation of the results for the different Hamiltonians is within the uncertainty due to the incompleteness of the basis set, which we estimate to be about 0.002 eV, see the discussion in Sec.~\ref{seq:discussion}.
\begin{table}[H]
\centering

\caption{EA for Og calculated with the FSCC-SD method exploiting various Hamiltonians.
The optimized basis set from the calculations with the $H_{\mathrm{GRECP}}$ Hamiltonian is used.
The electrons $6s6p5f6d7s7p$ are correlated.
AS includes the virtual DF orbitals with energies $\varepsilon^{\mathrm{DF}} < 40$ a.u..}
\label{table:2}
\begin{tabular}{l S[table-format=1.4]} 

\toprule
\multicolumn{1}{c}{Hamiltonian}  &
\multicolumn{1}{c}{$\mathrm{EA}$, eV} \\
\midrule
$H_{\mathrm{GRECP}}$ & 0.0725 \\
$H_{\mathrm{X2Cmmf}}$ & 0.0717 \\
$H_{\mathrm{DC}}$ &  0.0707 \\
\bottomrule

\end{tabular}
\end{table}
\par
The correction to EA associated with the Gaunt interaction, $\delta \epsilon^{\mathrm{G}}$, is calculated in accordance with Eq.~(\ref{eq:EA^G}) using the $H_{\mathrm{X2Cmmf}}$ Hamiltonian with the optimized basis set.
Table~\ref{table:3} presents the results of the calculations of the Gaunt interaction correction to EA.
It shows the dependence of the ground-state energy of the atom and EA calculated with and without the Gaunt interaction on the number of the correlated electrons and the number of the virtual DF orbitals included into AS.
The quantum numbers of the correlated electrons are presented in the first column, the maximum energy of the DF orbital included into the active space, $\varepsilon^{\mathrm{DF}}_{\mathrm{max}}$, is in the second column, the ground state energy of the atom calculated with and without $V^{\mathrm{G}}$ is given in the third and fourth columns, EA calculated with and without $V^{\mathrm{G}}$ operator is shown in the next two columns, and the correction to EA from the Gaunt interaction is given in the last column.
\par
Although the Gaunt interaction increases the ground state energy of the atom $E^{0h0p}$ by about $90$ a.u., it also shifts the energy of the anion for the same value.
Therefore, the resulting correction from the Gaunt interaction to EA, $\delta \epsilon^{\mathrm{G}}$, calculated with the $H_{\mathrm{X2Cmmf}}$ Hamiltonian at the FSCC-SD correlation level amounts to about $-0.0001$ eV.
This correction does not depend on the active-space size since the inclusion of the $2s...7p$ electrons into the correlation problem and adding the virtual states with energies up to the $6200$ a.u. into AS do not change the value of $\delta \epsilon^{\mathrm{G}}$.
Though, the energies of the atom and anion change by about $-3$ a.u. compared to the calculation where only the $5d...7p$ electrons are correlated and the virtual states with energies $\varepsilon^{\mathrm{DF}} < 40$ a.u. are included into AS.
The calculated correction $\delta \epsilon^{\mathrm{G}}$ is an order of magnitude smaller than the uncertainty associated with the basis set.
\par
Proceeding with the analysis, we estimate the correction to EA from highly excited virtual states and strongly bound core electrons.
The difference between EA, calculated with the $6s...7p$ electrons correlated and the virtual states with $\varepsilon^{\mathrm{DF}} < 40$ a.u. included into the active space, and EA, calculated with the $2s..7p$ electrons correlated and the virtual states with $\varepsilon^{\mathrm{DF}} < 6200$ a.u. included into AS, is about $0.0008$ eV.
The same difference where the second term is calculated with the $5d..7p$ electrons correlated and the virtual states with $\varepsilon^{\mathrm{DF}} < 80$ a.u. included into AS is $0.0002$ eV.
We conclude that the uncertainty of the EA value associated with the correlation correction from the $1s...4f$ electrons is about $0.0005$ eV which is three times less than the uncertainty due to the incomplete basis set.
\par
The correction to EA from the full iterative triple~(T) excitations, which is defined as $\delta \epsilon^{\mathrm{T}} = \epsilon^{\mathrm{SDT}}_{\mathrm{FSCC}} - \epsilon^{\mathrm{SD}}_{\mathrm{FSCC}}$, is evaluated using the EXP-T program \cite{2020_OleynichenkoA_Symmetry}.
Since the FSCC-SDT equations are much more complicated compared to the FSCC-SD ones, some additional reductions of the problem have been made.
Instead of constructing an optimized basis for solving the FSCC-SDT problem, we use the basis which was optimized for the FSCC-SD calculations.
We also employ the $H_{\mathrm{GRECP}}$ Hamiltonian while dealing with the T excitation correction.
We study the dependence of $\delta \epsilon^{\mathrm{T}}$ with respect to the AS size and our final result for the $\delta \epsilon^{\mathrm{T}}$ correction is~$0.008(3)$~eV.
\onecolumngrid\
\begin{table}[H]
\centering

\caption{The ground state energies $E^{0h0p}_{\mathrm{G}}$ and $E^{0h0p}$ calculated at the FSCC-SD level with and without the Gaunt interaction, respectively; the electron affinities $\epsilon_{\mathrm{X2Cmmf}}^{\mathrm{G}}$ and $\epsilon_{\mathrm{X2Cmmf}}$ evaluated with and without the Gaunt interaction, respectively; and the correction to EA from the Gaunt interaction, $\delta \epsilon^{\mathrm{G}}$, depending on the quantum numbers of the electrons included into the correlation problem and the maximum energy of the virtual DF orbitals,  $\varepsilon_{\mathrm{max}}^{\mathrm{DF}}$, included into AS.
The energies $E^{0h0p}_{\mathrm{G}}$, $E^{0h0p}$, and $\varepsilon_{\mathrm{max}}^{\mathrm{DF}}$ are in a.u., whereas the electron affinities $\epsilon_{\mathrm{X2Cmmf}}^{\mathrm{G}}$, $\epsilon_{\mathrm{X2Cmmf}}$, and $\delta \epsilon^{\mathrm{G}}$ are in eV.}
\label{table:3}
\begin{tabular}{
c
S[table-format=6]
S[table-format=-5.6]
S[table-format=-5.6]
S[table-format=1.8]
S[table-format=1.8]
S[table-format=-1.6]} 

\toprule
\multicolumn{1}{c}{Electrons} &
$\varepsilon^{\mathrm{DF}}_{\mathrm{max}}\textrm{,}\,\,\textrm{a.u.}$ &
\multicolumn{1}{c}{$E^{0h0p}_{\mathrm{G}}\textrm{,}\,\,\textrm{a.u.}$}  &
\multicolumn{1}{c}{$E^{0h0p}\textrm{,}\,\,\textrm{a.u.}$} &
\multicolumn{1}{c}{$\epsilon_{\mathrm{X2Cmmf}}^{\mathrm{G}}\textrm{,}\,\,\textrm{eV}$} &
\multicolumn{1}{c}{$\epsilon_{\mathrm{X2Cmmf}}\textrm{,}\,\,\textrm{eV}$} & 
$\delta \epsilon^{\mathrm{G}}\textrm{,}\,\,\textrm{eV}$  \\
\midrule
$6s...7p$& 40	  &  -54732.65683 & -54842.50513 &  0.07160 & 0.07165 & -0.00006 \\
$5d...7p$& 80    &  -54732.78035 & -54842.63163 &  0.07093 & 0.07099 & -0.00006 \\
$4f...7p$& 170	  &  -54733.16953 & -54843.02129 &  0.07068 & 0.07074 & -0.00006 \\
$4d...7p$& 270   &  -54733.53866 & -54843.39065 &  0.07089 & 0.07095 & -0.00006 \\
$4s...7p$& 500   &  -54733.68575 & -54843.53806 &  0.07070 & 0.07076 & -0.00006 \\
$3d...7p$& 980   &  -54734.29637 & -54844.14919 &  0.07083 & 0.07089 & -0.00006 \\
$3s...7p$& 2000  &  -54734.50493 & -54844.35852 &  0.07077 & 0.07083 & -0.00006 \\
$2s...7p$& 6200  &  -54735.05391 & -54844.90824 &  0.07076 & 0.07082 & -0.00006 \\
\bottomrule

\end{tabular}
\end{table}
\twocolumngrid\

\subsection{CI-DFS details}\label{subseq:methods:cidfs}
To obtain the bound-state solution for the $8s$ orbital at the DF level, we introduce the polarization potential $V^{\mathrm{pol}}_{8s}$ into the DF and DFS equations.
According to the generalized Koopmans theorem, the energies of the DF orbitals correspond to the EA values in the DF approximation taken with the opposite sign.
Therefore, first, we adjust the parameter $r_{\mathrm{cut}}$ in Eq.~(\ref{eq:V^pol}) to reproduce qualitatively the DF energy of $\varphi^{\mathrm{DF}}_{8s}$ at a given value of the parameter $\alpha_d=58.5$ a.u..
The employed value for the parameter $r_{\mathrm{cut}}$ is 3.3 a.u..
The value $\alpha_d=58.5(15)$ a.u. is calculated in the present work using the finite-field approach in the framework of the CC-SD method.
Our value of $\alpha_d$ is in good agreement with the results of Refs.~\cite{2016_DzubaV_PhysRevA, 2018_JerabekP_PhysRevLett}.
Having obtained the physically justified $\varphi^{\mathrm{DF}}_{8s}$ orbital and constructed the basis of virtual DFS orbitals, we proceed to the CI correlation problem.
\par
In the CI-DFS method, EA is defined according to Eq.~(\ref{eq:EA_CI}) as the energy difference of the two charge states of Og. 
The total energies of Og and Og$^-$ in absolute magnitude are larger than $50000$ a.u., though their difference is only a tenth of eV.
Varying the scheme of constructing the virtual orbitals or/and the many-electron basis for an individual state, we can obtain a lower total energy for this state. 
However, the difference between the two large energies can be unstable with respect to the parameters of the employed configuration spaces for Og and Og$^-$.
Therefore, we focus on constructing such a basis set that describes in a \textit{balanced} manner the difference of the energies, and we consider the stability of the results for EA as a primary criterion for the basis-set choice.
Having this remark kept in mind, we exploit the anionic set of the occupied and virtual orbitals for both Og and Og$^-$ calculations since it results in the more accurate cancellation of the core-core and core-valence correlation effects in the energy difference.
\par
Throughout the calculations, we consider the orbital $7p$, as well as $8s$ for the case of Og$^-$, as the active occupied ones.
We separate the contributions from the SD and partially from the T and quadruple (Q) excitations to EA and study them individually with respect to the basis set enlargement.
The contribution to EA from the SD excitations is defined as
\begin{equation}\label{eq:EA_CI_SD}
    \epsilon^{\mathrm{SD}}_{\mathrm{CI}} = E^{\mathrm{SD}}_{\mathrm{CI}}(\mathrm{Og}) - E^{\mathrm{SD}}_{\mathrm{CI}}(\mathrm{Og}^-).
\end{equation}
Not all the T excitations were taken into account, but only the most important part of them for the problem under consideration, which corresponds to the simultaneous D excitations from $7p$ and S excitations from $8s$ to the virtual orbitals.
We denote them as [T].
In this particular case, when we are dealing with the charged states of Og and Og$^-$, the inclusion of these excitations can partly restore the size-extensivity property, which is absent in the truncated CI-SD method.
The contribution from the [T] excitations to EA is then
\begin{equation}
    \epsilon^{\mathrm{[T]}}_{\mathrm{CI}} = E^{\mathrm{SD[T]}}_{\mathrm{CI}}(\mathrm{Og}) - E^{\mathrm{SD[T]}}_{\mathrm{CI}}(\mathrm{Og}^-) - \epsilon^{\mathrm{SD}}_{\mathrm{CI}}.
\end{equation}
The quadruple excitations, denoted as [Q], are also included in an analogues way as [T]
and correspond to the simultaneous triple excitations from $7p$ and S excitations from $8s$.
The contribution from the [Q] excitations to EA reads as 
\begin{equation}\label{eq:EA_CI_Q}
    \epsilon^{\mathrm{[Q]}}_{\mathrm{CI}} = E^{\mathrm{SDT[Q]}}_{\mathrm{CI}}(\mathrm{Og}) - E^{\mathrm{SDT[Q]}}_{\mathrm{CI}}(\mathrm{Og}^-) - \epsilon^{\mathrm{[T]}}_{\mathrm{CI}} - \epsilon^{\mathrm{SD}}_{\mathrm{CI}}.
\end{equation}
\par
In Table~\ref{table:4}, the results for the contributions from the excitations of different types to the $\epsilon_{\mathrm{CI}}$ value are collected.
The uncertainty has a pure numerical origin and is due to the convergence of the results with respect to the number of virtual orbitals.
The contributions $\epsilon^{\mathrm{[T]}}_{\mathrm{CI}}$ and $\epsilon^{\mathrm{[Q]}}_{\mathrm{CI}}$ are evaluated with a smaller number of virtual orbitals compared to the $\epsilon^{\mathrm{SD}}_{\mathrm{CI}}$ value.
Meanwhile, these contributions turn out to be several times more important than the SD contributions.
The main source of the uncertainty arises from the evaluation of the [Q] excitations.
\begin{table}[htbp]
\centering

\caption{The contributions to $\epsilon_{\mathrm{CI}}$ from the SD, [T] and [Q] excitations, in eV. 
The [T] excitations are the dominant part of all T ones and correspond to the simultaneous D excitations from $7p$ and S excitations from $8s$. 
On the other hand, the [Q] excitations are the dominant part of all Q ones and correspond to the simultaneous T excitations from $7p$ and S excitations from $8s$.
The last row shows the sum of all considered contributions to $\epsilon_{\mathrm{CI}}$.}
\label{table:4}
\begin{tabular}{c S[table-format=2.3(1)]}

\toprule
\multicolumn{1}{c}{Contribution}  &
\multicolumn{1}{c}{Value, eV}  \\
\midrule
$\epsilon_{\mathrm{CI}}$  & 0.008(2) \\
$\epsilon_{\mathrm{CI}}^{\mathrm{[T]}}$  & 0.046(3) \\
$\epsilon_{\mathrm{CI}}^{\mathrm{[Q]}}$  & 0.016(10) \\
\vspace{+0.25 pt}
$\epsilon_{\mathrm{CI}}^{}$  & 0.070(10) \\
\bottomrule

\end{tabular}
\end{table}
\par
Within the framework of the CI-DFS method we have also calculated the contributions to EA from the Gaunt and retardation corrections defined by Eqs. (\ref{eq:V^Gaunt}) and (\ref{eq:V^Ret}), respectively.
The Gaunt correction coincides with the corresponding FSCC result presented in Table \ref{table:3}.
The evaluated retardation effect amounts to about $-0.0003$ eV.
The QED correction to EA is calculated according to Eq.~(\ref{eq:dEA_QED}).
As in the case of the calculations of $\epsilon_{\mathrm{CI}}$, we separated the SD, [T], and [Q] excitation contributions to $\epsilon^{\mathrm{QED}}_{\mathrm{CI}}$ and studied the convergence of the result with respect to the number of virtual orbitals, see the related discussion in Sec.~\ref{seq:discussion}.
\section{Results and Discussion}\label{seq:discussion}
Using the optimized basis set, obtained in accordance with the procedure described in App.~\ref{seq:appendix_a}, we perform the FSCC-SD calculations of EA for Og with the $H_{\mathrm{DC}}$ Hamiltonian.
In these calculations, the $5s...7p$ electrons have been correlated and the virtual states with the energies $\varepsilon^{\mathrm{DF}} < 80 $ a.u. have been included into AS.
The obtained value for EA is $\epsilon_{\mathrm{FSCC}}^{\mathrm{SD}}=0.070(2)$ eV.
The correction from the T excitations is calculated with a smaller basis using the rigorous solution of the FSCC-SDT equations for the sectors $0h0p$ and $0h1p$ by means of the EXP-T program \cite{2020_OleynichenkoA_Symmetry}.
As a result, this correction is obtained to be $\delta \epsilon_{\mathrm{FSCC}}^{\mathrm{T}} = 0.008(3)$ eV.
Thus, we consider our final value for EA, evaluated with the FSCC method, to be $\epsilon_{\mathrm{FSCC}}=0.078(4)$ eV, with the basis-set incompleteness providing the main source of the uncertainty.
\par
Using the CI-DFS method in accordance with the procedure discussed in Sec.~\ref{subseq:methods:cidfs}, we have obtained the value $\mathrm{\epsilon_{CI}}=0.070(10)$~eV.
However, as compared to the FSCC result, in our CI-DFS calculations the excitations beyond SD, [T], and [Q] are absent and only the $7p8s$ electrons are correlated.
In contrast, in the FSCC calculations, the $5d6s6p5f6d7s7p8s$ electrons were correlated, and more types of the excitations were included due to the CC ansatz.
Having studied the correction from the inclusion of the $5d6s6p5f6d7s$ electrons into the FSCC-SD correlation problem, we found that this correction does not exceed $0.002$~eV and, thereby, is covered by the uncertainty of $\mathrm{\epsilon_{CI}}$.
Therefore, we believe that the main reason for the deviation of our CI-DFS results from the FSCC ones is the absence of the fully T and higher-order excitations in the CI-DFS calculations.
Although the $\mathrm{\epsilon_{CI}}$ and $\mathrm{\epsilon_{FSCC}}$ values, obtained in the present work within the conceptually different methods, are in reasonable agreement with each other, we consider our FSCC results to be more reliable.
\par
The QED correction is obtained using the model QED operator \cite{2013_ShabaevV_PhysRevA, 2015_ShabaevV_CompPhysComm} and is extracted from our CI-DFS calculations.
The operator~$V^{\mathrm{QED}}_{\mathrm{mod}}$ was included into the DF and DFS equations, which define the occupied and virtual orbitals employed in the CI procedure, as well as into the $H_{\mathrm{DCB}}$ Hamiltonian.
The resulted value is $\delta \epsilon^{\mathrm{QED}} = -0.002(1)$ eV.
It was found that the correction $\delta \epsilon^{\mathrm{QED}}$ is more stable than the individual contributions $\epsilon_{\mathrm{CI}}^{\mathrm{QED}}$ and $\epsilon_{\mathrm{CI}}$.
Therefore, the numerical uncertainty associated with the convergence of the correction $\delta \epsilon^{\mathrm{QED}}$ with respect to the AS enlargement is also smaller than that for the individual $\epsilon^{\mathrm{QED}}_{\mathrm{CI}}$ and $\epsilon_{\mathrm{CI}}$ terms.
In addition, the dominant contribution to the $\delta \epsilon_{\mathrm{CI}}^{\mathrm{QED}}$ correction comes from the SD excitations, but not from the [T] ones, as it takes place for the individual terms.
The uncertainty of $\delta \epsilon^{\mathrm{QED}}$ given above includes not only the part associated with the convergence of the result with respect to the number of the virtual orbitals and the size of AS but also a conservative estimate of the higher-order QED effects, which are beyond the model QED operator approach.
The same value for the QED correction to EA was obtained within the framework of the coupled cluster theory using the very recent implementation of the model QED operator for calculations of electronic energies in molecular systems~\cite{Skripnikov:2021}.
The results for EA of Og obtained in the present work and the corresponding results from Refs.~\cite{1996_EliavE_PhysRevLett, 2003_GoidenkoI_PhysRevA, 2018_LackenbyB_PhysRevA98_Og} are presented in Table~\ref{table:5}.
\begin{table}[htbp]
\centering

\begin{threeparttable}

\caption{Comparison of EA for Og and the QED contribution to EA calculated with the CI-DFS and FSCC methods in the present work with results of Refs. \cite{1996_EliavE_PhysRevLett, 2003_GoidenkoI_PhysRevA, 2018_LackenbyB_PhysRevA98_Og}, in eV.
In Ref.~\cite{2018_LackenbyB_PhysRevA98_Og}, a combination of CI with the many-body perturbation theory is used.}
\label{table:5}

\begin{tabular}{l S[table-align-text-post = false, table-number-alignment=left, table-format=1.3(2)] S[table-format=1.3(2)] S[table-format=-1.4(1)] S[table-format=1.3(1)]}

\toprule
\multicolumn{1}{c}{Reference}  &
\multicolumn{1}{c}{$\epsilon_{\mathrm{CI}}$}  & 
\multicolumn{1}{c}{$\epsilon_{\mathrm{FSCC}}$}  & 
\multicolumn{1}{c}{$\delta \epsilon^{\mathrm{QED}}$} &
\multicolumn{1}{c}{$\epsilon^{\mathrm{Total}}$} \\
\midrule
Present Work & 0.070(10) & 0.078(4) &  -0.002(1) & 0.076(4)\\
Eliav \textit{et al.} \cite{1996_EliavE_PhysRevLett} & & 0.056(10)  &  & \\
Goidenko \textit{et al.} \cite{2003_GoidenkoI_PhysRevA} & & 0.064(2) & -0.0059(5) & 0.058(3) \\
Lackenby \textit{et al.} \cite{2018_LackenbyB_PhysRevA98_Og} &  &  & & 0.096 \\
\bottomrule
\end{tabular}


\end{threeparttable}
\end{table}
\par
The main difference between our results and those from Refs.~\cite{1996_EliavE_PhysRevLett, 2003_GoidenkoI_PhysRevA} is due to the choice of the basis set.
In Refs.~\cite{1996_EliavE_PhysRevLett, 2003_GoidenkoI_PhysRevA}, a universal (with the same parameters of the basis functions for all elements) Gaussian basis set was exploited, which turned out to be rather inappropriate for this particular problem.
Moreover, in contrast to Refs.~\cite{1996_EliavE_PhysRevLett, 2003_GoidenkoI_PhysRevA}, we rigorously evaluated the correction from the triple excitations.
Overall, our FSCC-SD value, which is $0.070(2)$ eV, is in reasonable agreement with the result of Ref.~\cite{2003_GoidenkoI_PhysRevA}.
However, the obtained QED correction is three times smaller than that reported in Ref.~\cite{2003_GoidenkoI_PhysRevA}.
\par
To identify the origin of the latter discrepancy, we also estimate $\delta\epsilon^{\mathrm{QED}}$ as the expectation value of the operator $V^{\mathrm{QED}}_{\mathrm{mod}}$ for the $8s$ spinor associated with the attached electron.
This spinor should be constructed with a proper account of the electron-electron correlation (let us recall that the corresponding DF spinor would correspond to a scattered electron state, provided that the basis is flexible enough).
The simplest approximation for the bound-state $8s$ spinor is obtained by converting the lowest-energy eigenvector of the FSCC effective Hamiltonian in the $0h1p$ sector into the single determinant, $P_{\mathrm{as}}\tilde{\varphi}_{8s}\Phi_0$, where $P_{\mathrm{as}}$ stands for the antisymmetrizer, $\Phi_0$ denotes the Fermi vacuum state, and $\tilde{\varphi}_{8s}$ is a linear combination of ``valence'' DF $s$-spinors.
A better approximation can be obtained as
\begin{equation}
    N(1+T^{0h1p}_1)\tilde{\varphi}_{8s},
 \label{machin}
\end{equation}
where $T^{0h1p}_1$ is the single-excitation part of the cluster operator in the $0h1p$ sector and $N$ stands for the normalizing factor.
The expectation value of $V^{\mathrm{QED}}_{\mathrm{mod}}$ for the spinor (\ref{machin}) tends to $-0.002$ eV, thus being consistent with our $\delta\epsilon^{\mathrm{QED}}$.
In contrast, evaluating its counterpart only with $\tilde{\varphi}_{8s}$ for restricted MS, in other word, neglecting the contributions of higher-energy DF $s$-spinors to the shape of the $8s$ natural spinor, one can readily reproduce the results of Ref.~\cite{2003_GoidenkoI_PhysRevA}.
\par
Finally, we have also separately compared the self-energy (SE) and vacuum-polarization (VP) contributions with the corresponding corrections presented in Ref.~\cite{2003_GoidenkoI_PhysRevA}.
For this purpose, we evaluated the expectation value of the SE and VP parts of the model QED operator with the function $\tilde{\varphi}_{8s}$.
The VP correction was additionally decomposed into a sum of the Uehling (Ue) and Wichmann-Kroll (WK) terms, since in Ref.~\cite{2003_GoidenkoI_PhysRevA} only the Ue potential was considered.
The expectation values of the SE and Ue parts of the operator $V^{\mathrm{QED}}_{\mathrm{mod}}$ have been compared with the related values from Ref.~\cite{2003_GoidenkoI_PhysRevA}.
The comparison shows a similar trend: when a small MS is considered and the second term in Eq.~(\ref{machin}) is absent, then our Ue and SE expectation values are close to the ones reported in Ref.~\cite{2003_GoidenkoI_PhysRevA}.
Nevertheless, when both terms in Eq.~(\ref{machin}) are considered, our expectation values for SE and VP become several times smaller, resulting in $-0.002$ eV for the sum of them.
Therefore, we guess that the primary source of the discrepancy between our QED correction and that given in Ref.~\cite{2003_GoidenkoI_PhysRevA} is caused by a small MS employed in that work. 
\par
Our total EA value also deviates from that obtained in Ref.~\cite{2018_LackenbyB_PhysRevA98_Og}.
The reason for this deviation is not clear to us since the work \cite{2018_LackenbyB_PhysRevA98_Og} does not contain any discussion of the uncertainty of the EA calculations.
\section{Summary}\label{seq:conclusion}
In the present work, the electron affinity of Og is calculated using two different methods: FSCC which is implemented in the DIRAC \cite{2020_SaueT_JChemPhys} and EXP-T packages \cite{2020_OleynichenkoA_Symmetry} and CI which is implemented in the CI-DFS program \cite{2003_TupitsynI_PhysRevA, 2005_TupitsynI_PhysRevA, 2018_TupitsynI_PhysRevA}.
The FSCC equations with single, double, and triple excitations are solved in the specially optimized basis set with the help of the EXP-T package, and the correction from the triple excitations to EA of Og is rigorously evaluated.
This correction turns out to be significant and amounts to 11\% of the FSCC-SD value.
\par
Both FSCC and CI-DFS methods exploited in the present work yield the results which are in reasonable agreement with each other.
The QED correction to EA of Og is calculated employing the QEDMOD operator \cite{2013_ShabaevV_PhysRevA, 2015_ShabaevV_CompPhysComm} within the CI-DFS method.
This correction turns out to be three times smaller than the previous value from Ref.~\cite{2003_GoidenkoI_PhysRevA} and comparable with the uncertainty associated with the electronic-correlation effects.
The total value of EA for Og is obtained to amount to 0.076(4) eV, which includes the correlation effects calculated by the FSCC-SDT method and the QED correction evaluated by means of the CI-DFS method combined with the QEDMOD operator.
\section{Acknowledgements}
The work is supported by the Ministry of Science and Higher Education of the Russian Federation within the~Grant~No.~075-10-2020-117.

\bibliographystyle{apsrev4-2}
\bibliography{main}

\begin{thebibliography}{69}%
\makeatletter
\providecommand \@ifxundefined [1]{%
 \@ifx{#1\undefined}
}%
\providecommand \@ifnum [1]{%
 \ifnum #1\expandafter \@firstoftwo
 \else \expandafter \@secondoftwo
 \fi
}%
\providecommand \@ifx [1]{%
 \ifx #1\expandafter \@firstoftwo
 \else \expandafter \@secondoftwo
 \fi
}%
\providecommand \natexlab [1]{#1}%
\providecommand \enquote  [1]{``#1''}%
\providecommand \bibnamefont  [1]{#1}%
\providecommand \bibfnamefont [1]{#1}%
\providecommand \citenamefont [1]{#1}%
\providecommand \href@noop [0]{\@secondoftwo}%
\providecommand \href [0]{\begingroup \@sanitize@url \@href}%
\providecommand \@href[1]{\@@startlink{#1}\@@href}%
\providecommand \@@href[1]{\endgroup#1\@@endlink}%
\providecommand \@sanitize@url [0]{\catcode `\\12\catcode `\$12\catcode
  `\&12\catcode `\#12\catcode `\^12\catcode `\_12\catcode `\%12\relax}%
\providecommand \@@startlink[1]{}%
\providecommand \@@endlink[0]{}%
\providecommand \url  [0]{\begingroup\@sanitize@url \@url }%
\providecommand \@url [1]{\endgroup\@href {#1}{\urlprefix }}%
\providecommand \urlprefix  [0]{URL }%
\providecommand \Eprint [0]{\href }%
\providecommand \doibase [0]{https://doi.org/}%
\providecommand \selectlanguage [0]{\@gobble}%
\providecommand \bibinfo  [0]{\@secondoftwo}%
\providecommand \bibfield  [0]{\@secondoftwo}%
\providecommand \translation [1]{[#1]}%
\providecommand \BibitemOpen [0]{}%
\providecommand \bibitemStop [0]{}%
\providecommand \bibitemNoStop [0]{.\EOS\space}%
\providecommand \EOS [0]{\spacefactor3000\relax}%
\providecommand \BibitemShut  [1]{\csname bibitem#1\endcsname}%
\let\auto@bib@innerbib\@empty
\bibitem [{\citenamefont {Oganessian}\ \emph {et~al.}(2006)\citenamefont
  {Oganessian}, \citenamefont {Utyonkov}, \citenamefont {Lobanov},
  \citenamefont {Abdullin}, \citenamefont {Polyakov}, \citenamefont {Sagaidak},
  \citenamefont {Shirokovsky}, \citenamefont {Tsyganov}, \citenamefont
  {Voinov}, \citenamefont {Gulbekian}, \citenamefont {Bogomolov}, \citenamefont
  {Gikal}, \citenamefont {Mezentsev}, \citenamefont {Iliev}, \citenamefont
  {Subbotin}, \citenamefont {Sukhov}, \citenamefont {Subotic}, \citenamefont
  {Zagrebaev}, \citenamefont {Vostokin}, \citenamefont {Itkis}, \citenamefont
  {Moody}, \citenamefont {Patin}, \citenamefont {Shaughnessy}, \citenamefont
  {Stoyer}, \citenamefont {Stoyer}, \citenamefont {Wilk}, \citenamefont
  {Kenneally}, \citenamefont {Landrum}, \citenamefont {Wild},\ and\
  \citenamefont {Lougheed}}]{2006_OganessianY_PhysRevC}%
  \BibitemOpen
  \bibfield  {author} {\bibinfo {author} {\bibfnamefont {Y.~T.}\ \bibnamefont
  {Oganessian}}, \bibinfo {author} {\bibfnamefont {V.~K.}\ \bibnamefont
  {Utyonkov}}, \bibinfo {author} {\bibfnamefont {Y.~V.}\ \bibnamefont
  {Lobanov}}, \bibinfo {author} {\bibfnamefont {F.~S.}\ \bibnamefont
  {Abdullin}}, \bibinfo {author} {\bibfnamefont {A.~N.}\ \bibnamefont
  {Polyakov}}, \bibinfo {author} {\bibfnamefont {R.~N.}\ \bibnamefont
  {Sagaidak}}, \bibinfo {author} {\bibfnamefont {I.~V.}\ \bibnamefont
  {Shirokovsky}}, \bibinfo {author} {\bibfnamefont {Y.~S.}\ \bibnamefont
  {Tsyganov}}, \bibinfo {author} {\bibfnamefont {A.~A.}\ \bibnamefont
  {Voinov}}, \bibinfo {author} {\bibfnamefont {G.~G.}\ \bibnamefont
  {Gulbekian}}, \bibinfo {author} {\bibfnamefont {S.~L.}\ \bibnamefont
  {Bogomolov}}, \bibinfo {author} {\bibfnamefont {B.~N.}\ \bibnamefont
  {Gikal}}, \bibinfo {author} {\bibfnamefont {A.~N.}\ \bibnamefont
  {Mezentsev}}, \bibinfo {author} {\bibfnamefont {S.}~\bibnamefont {Iliev}},
  \bibinfo {author} {\bibfnamefont {V.~G.}\ \bibnamefont {Subbotin}}, \bibinfo
  {author} {\bibfnamefont {A.~M.}\ \bibnamefont {Sukhov}}, \bibinfo {author}
  {\bibfnamefont {K.}~\bibnamefont {Subotic}}, \bibinfo {author} {\bibfnamefont
  {V.~I.}\ \bibnamefont {Zagrebaev}}, \bibinfo {author} {\bibfnamefont {G.~K.}\
  \bibnamefont {Vostokin}}, \bibinfo {author} {\bibfnamefont {M.~G.}\
  \bibnamefont {Itkis}}, \bibinfo {author} {\bibfnamefont {K.~J.}\ \bibnamefont
  {Moody}}, \bibinfo {author} {\bibfnamefont {J.~B.}\ \bibnamefont {Patin}},
  \bibinfo {author} {\bibfnamefont {D.~A.}\ \bibnamefont {Shaughnessy}},
  \bibinfo {author} {\bibfnamefont {M.~A.}\ \bibnamefont {Stoyer}}, \bibinfo
  {author} {\bibfnamefont {N.~J.}\ \bibnamefont {Stoyer}}, \bibinfo {author}
  {\bibfnamefont {P.~A.}\ \bibnamefont {Wilk}}, \bibinfo {author}
  {\bibfnamefont {J.~M.}\ \bibnamefont {Kenneally}}, \bibinfo {author}
  {\bibfnamefont {J.~H.}\ \bibnamefont {Landrum}}, \bibinfo {author}
  {\bibfnamefont {J.~F.}\ \bibnamefont {Wild}},\ and\ \bibinfo {author}
  {\bibfnamefont {R.~W.}\ \bibnamefont {Lougheed}},\ }\href
  {https://doi.org/10.1103/PhysRevC.74.044602} {\bibfield  {journal} {\bibinfo
  {journal} {Physical Review C}\ }\textbf {\bibinfo {volume} {74}},\ \bibinfo
  {pages} {044602} (\bibinfo {year} {2006})}\BibitemShut {NoStop}%
\bibitem [{\citenamefont {Oganessian}\ \emph {et~al.}(2012)\citenamefont
  {Oganessian}, \citenamefont {Abdullin}, \citenamefont {Alexander},
  \citenamefont {Binder}, \citenamefont {Boll}, \citenamefont {Dmitriev},
  \citenamefont {Ezold}, \citenamefont {Felker}, \citenamefont {Gostic},
  \citenamefont {Grzywacz}, \citenamefont {Hamilton}, \citenamefont
  {Henderson}, \citenamefont {Itkis}, \citenamefont {Miernik}, \citenamefont
  {Miller}, \citenamefont {Moody}, \citenamefont {Polyakov}, \citenamefont
  {Ramayya}, \citenamefont {Roberto}, \citenamefont {Ryabinin}, \citenamefont
  {Rykaczewski}, \citenamefont {Sagaidak}, \citenamefont {Shaughnessy},
  \citenamefont {Shirokovsky}, \citenamefont {Shumeiko}, \citenamefont
  {Stoyer}, \citenamefont {Stoyer}, \citenamefont {Subbotin}, \citenamefont
  {Sukhov}, \citenamefont {Tsyganov}, \citenamefont {Utyonkov}, \citenamefont
  {Voinov},\ and\ \citenamefont {Vostokin}}]{2012_OganessianY_PhysRevLett}%
  \BibitemOpen
  \bibfield  {author} {\bibinfo {author} {\bibfnamefont {Y.~T.}\ \bibnamefont
  {Oganessian}}, \bibinfo {author} {\bibfnamefont {F.~S.}\ \bibnamefont
  {Abdullin}}, \bibinfo {author} {\bibfnamefont {C.}~\bibnamefont {Alexander}},
  \bibinfo {author} {\bibfnamefont {J.}~\bibnamefont {Binder}}, \bibinfo
  {author} {\bibfnamefont {R.~A.}\ \bibnamefont {Boll}}, \bibinfo {author}
  {\bibfnamefont {S.~N.}\ \bibnamefont {Dmitriev}}, \bibinfo {author}
  {\bibfnamefont {J.}~\bibnamefont {Ezold}}, \bibinfo {author} {\bibfnamefont
  {K.}~\bibnamefont {Felker}}, \bibinfo {author} {\bibfnamefont {J.~M.}\
  \bibnamefont {Gostic}}, \bibinfo {author} {\bibfnamefont {R.~K.}\
  \bibnamefont {Grzywacz}}, \bibinfo {author} {\bibfnamefont {J.~H.}\
  \bibnamefont {Hamilton}}, \bibinfo {author} {\bibfnamefont {R.~A.}\
  \bibnamefont {Henderson}}, \bibinfo {author} {\bibfnamefont {M.~G.}\
  \bibnamefont {Itkis}}, \bibinfo {author} {\bibfnamefont {K.}~\bibnamefont
  {Miernik}}, \bibinfo {author} {\bibfnamefont {D.}~\bibnamefont {Miller}},
  \bibinfo {author} {\bibfnamefont {K.~J.}\ \bibnamefont {Moody}}, \bibinfo
  {author} {\bibfnamefont {A.~N.}\ \bibnamefont {Polyakov}}, \bibinfo {author}
  {\bibfnamefont {A.~V.}\ \bibnamefont {Ramayya}}, \bibinfo {author}
  {\bibfnamefont {J.~B.}\ \bibnamefont {Roberto}}, \bibinfo {author}
  {\bibfnamefont {M.~A.}\ \bibnamefont {Ryabinin}}, \bibinfo {author}
  {\bibfnamefont {K.~P.}\ \bibnamefont {Rykaczewski}}, \bibinfo {author}
  {\bibfnamefont {R.~N.}\ \bibnamefont {Sagaidak}}, \bibinfo {author}
  {\bibfnamefont {D.~A.}\ \bibnamefont {Shaughnessy}}, \bibinfo {author}
  {\bibfnamefont {I.~V.}\ \bibnamefont {Shirokovsky}}, \bibinfo {author}
  {\bibfnamefont {M.~V.}\ \bibnamefont {Shumeiko}}, \bibinfo {author}
  {\bibfnamefont {M.~A.}\ \bibnamefont {Stoyer}}, \bibinfo {author}
  {\bibfnamefont {N.~J.}\ \bibnamefont {Stoyer}}, \bibinfo {author}
  {\bibfnamefont {V.~G.}\ \bibnamefont {Subbotin}}, \bibinfo {author}
  {\bibfnamefont {A.~M.}\ \bibnamefont {Sukhov}}, \bibinfo {author}
  {\bibfnamefont {Y.~S.}\ \bibnamefont {Tsyganov}}, \bibinfo {author}
  {\bibfnamefont {V.~K.}\ \bibnamefont {Utyonkov}}, \bibinfo {author}
  {\bibfnamefont {A.~A.}\ \bibnamefont {Voinov}},\ and\ \bibinfo {author}
  {\bibfnamefont {G.~K.}\ \bibnamefont {Vostokin}},\ }\href
  {https://doi.org/10.1103/PhysRevLett.109.162501} {\bibfield  {journal}
  {\bibinfo  {journal} {Physical Review Letters}\ }\textbf {\bibinfo {volume}
  {109}},\ \bibinfo {pages} {162501} (\bibinfo {year} {2012})}\BibitemShut
  {NoStop}%
\bibitem [{\citenamefont {Sato}\ \emph {et~al.}(2015)\citenamefont {Sato},
  \citenamefont {Asai}, \citenamefont {Borschevsky}, \citenamefont {Stora},
  \citenamefont {Sato}, \citenamefont {Kaneya}, \citenamefont {Tsukada},
  \citenamefont {D{\"u}llmann}, \citenamefont {Eberhardt}, \citenamefont
  {Eliav}, \citenamefont {Ichikawa}, \citenamefont {Kaldor}, \citenamefont
  {Kratz}, \citenamefont {Miyashita}, \citenamefont {Nagame}, \citenamefont
  {Ooe}, \citenamefont {Osa}, \citenamefont {Renisch}, \citenamefont {Runke},
  \citenamefont {Sch{\"a}del}, \citenamefont {{Th{\"o}rle-Pospiech}},
  \citenamefont {Toyoshima},\ and\ \citenamefont
  {Trautmann}}]{2015_SatoT_Nature}%
  \BibitemOpen
  \bibfield  {author} {\bibinfo {author} {\bibfnamefont {T.~K.}\ \bibnamefont
  {Sato}}, \bibinfo {author} {\bibfnamefont {M.}~\bibnamefont {Asai}}, \bibinfo
  {author} {\bibfnamefont {A.}~\bibnamefont {Borschevsky}}, \bibinfo {author}
  {\bibfnamefont {T.}~\bibnamefont {Stora}}, \bibinfo {author} {\bibfnamefont
  {N.}~\bibnamefont {Sato}}, \bibinfo {author} {\bibfnamefont {Y.}~\bibnamefont
  {Kaneya}}, \bibinfo {author} {\bibfnamefont {K.}~\bibnamefont {Tsukada}},
  \bibinfo {author} {\bibfnamefont {C.~E.}\ \bibnamefont {D{\"u}llmann}},
  \bibinfo {author} {\bibfnamefont {K.}~\bibnamefont {Eberhardt}}, \bibinfo
  {author} {\bibfnamefont {E.}~\bibnamefont {Eliav}}, \bibinfo {author}
  {\bibfnamefont {S.}~\bibnamefont {Ichikawa}}, \bibinfo {author}
  {\bibfnamefont {U.}~\bibnamefont {Kaldor}}, \bibinfo {author} {\bibfnamefont
  {J.~V.}\ \bibnamefont {Kratz}}, \bibinfo {author} {\bibfnamefont
  {S.}~\bibnamefont {Miyashita}}, \bibinfo {author} {\bibfnamefont
  {Y.}~\bibnamefont {Nagame}}, \bibinfo {author} {\bibfnamefont
  {K.}~\bibnamefont {Ooe}}, \bibinfo {author} {\bibfnamefont {A.}~\bibnamefont
  {Osa}}, \bibinfo {author} {\bibfnamefont {D.}~\bibnamefont {Renisch}},
  \bibinfo {author} {\bibfnamefont {J.}~\bibnamefont {Runke}}, \bibinfo
  {author} {\bibfnamefont {M.}~\bibnamefont {Sch{\"a}del}}, \bibinfo {author}
  {\bibfnamefont {P.}~\bibnamefont {{Th{\"o}rle-Pospiech}}}, \bibinfo {author}
  {\bibfnamefont {A.}~\bibnamefont {Toyoshima}},\ and\ \bibinfo {author}
  {\bibfnamefont {N.}~\bibnamefont {Trautmann}},\ }\href
  {https://doi.org/10.1038/nature14342} {\bibfield  {journal} {\bibinfo
  {journal} {Nature}\ }\textbf {\bibinfo {volume} {520}},\ \bibinfo {pages}
  {209} (\bibinfo {year} {2015})}\BibitemShut {NoStop}%
\bibitem [{\citenamefont {Chhetri}\ \emph {et~al.}(2018)\citenamefont
  {Chhetri}, \citenamefont {Ackermann}, \citenamefont {Backe}, \citenamefont
  {Block}, \citenamefont {Cheal}, \citenamefont {Droese}, \citenamefont
  {D{\"u}llmann}, \citenamefont {Even}, \citenamefont {Ferrer}, \citenamefont
  {Giacoppo}, \citenamefont {G{\"o}tz}, \citenamefont {He{\ss}berger},
  \citenamefont {Huyse}, \citenamefont {Kaleja}, \citenamefont {Khuyagbaatar},
  \citenamefont {Kunz}, \citenamefont {Laatiaoui}, \citenamefont
  {Lautenschl{\"a}ger}, \citenamefont {Lauth}, \citenamefont {Lecesne},
  \citenamefont {Lens}, \citenamefont {Minaya~Ramirez}, \citenamefont {Mistry},
  \citenamefont {Raeder}, \citenamefont {Van~Duppen}, \citenamefont {Walther},
  \citenamefont {Yakushev},\ and\ \citenamefont
  {Zhang}}]{2018_ChhetriP_PhysRevLett}%
  \BibitemOpen
  \bibfield  {author} {\bibinfo {author} {\bibfnamefont {P.}~\bibnamefont
  {Chhetri}}, \bibinfo {author} {\bibfnamefont {D.}~\bibnamefont {Ackermann}},
  \bibinfo {author} {\bibfnamefont {H.}~\bibnamefont {Backe}}, \bibinfo
  {author} {\bibfnamefont {M.}~\bibnamefont {Block}}, \bibinfo {author}
  {\bibfnamefont {B.}~\bibnamefont {Cheal}}, \bibinfo {author} {\bibfnamefont
  {C.}~\bibnamefont {Droese}}, \bibinfo {author} {\bibfnamefont {C.~E.}\
  \bibnamefont {D{\"u}llmann}}, \bibinfo {author} {\bibfnamefont
  {J.}~\bibnamefont {Even}}, \bibinfo {author} {\bibfnamefont {R.}~\bibnamefont
  {Ferrer}}, \bibinfo {author} {\bibfnamefont {F.}~\bibnamefont {Giacoppo}},
  \bibinfo {author} {\bibfnamefont {S.}~\bibnamefont {G{\"o}tz}}, \bibinfo
  {author} {\bibfnamefont {F.~P.}\ \bibnamefont {He{\ss}berger}}, \bibinfo
  {author} {\bibfnamefont {M.}~\bibnamefont {Huyse}}, \bibinfo {author}
  {\bibfnamefont {O.}~\bibnamefont {Kaleja}}, \bibinfo {author} {\bibfnamefont
  {J.}~\bibnamefont {Khuyagbaatar}}, \bibinfo {author} {\bibfnamefont
  {P.}~\bibnamefont {Kunz}}, \bibinfo {author} {\bibfnamefont {M.}~\bibnamefont
  {Laatiaoui}}, \bibinfo {author} {\bibfnamefont {F.}~\bibnamefont
  {Lautenschl{\"a}ger}}, \bibinfo {author} {\bibfnamefont {W.}~\bibnamefont
  {Lauth}}, \bibinfo {author} {\bibfnamefont {N.}~\bibnamefont {Lecesne}},
  \bibinfo {author} {\bibfnamefont {L.}~\bibnamefont {Lens}}, \bibinfo {author}
  {\bibfnamefont {E.}~\bibnamefont {Minaya~Ramirez}}, \bibinfo {author}
  {\bibfnamefont {A.~K.}\ \bibnamefont {Mistry}}, \bibinfo {author}
  {\bibfnamefont {S.}~\bibnamefont {Raeder}}, \bibinfo {author} {\bibfnamefont
  {P.}~\bibnamefont {Van~Duppen}}, \bibinfo {author} {\bibfnamefont
  {T.}~\bibnamefont {Walther}}, \bibinfo {author} {\bibfnamefont
  {A.}~\bibnamefont {Yakushev}},\ and\ \bibinfo {author} {\bibfnamefont
  {Z.}~\bibnamefont {Zhang}},\ }\href
  {https://doi.org/10.1103/PhysRevLett.120.263003} {\bibfield  {journal}
  {\bibinfo  {journal} {Physical Review Letters}\ }\textbf {\bibinfo {volume}
  {120}},\ \bibinfo {pages} {263003} (\bibinfo {year} {2018})}\BibitemShut
  {NoStop}%
\bibitem [{\citenamefont {Raeder}\ \emph {et~al.}(2018)\citenamefont {Raeder},
  \citenamefont {Ackermann}, \citenamefont {Backe}, \citenamefont {Beerwerth},
  \citenamefont {Berengut}, \citenamefont {Block}, \citenamefont {Borschevsky},
  \citenamefont {Cheal}, \citenamefont {Chhetri}, \citenamefont {D{\"u}llmann},
  \citenamefont {Dzuba}, \citenamefont {Eliav}, \citenamefont {Even},
  \citenamefont {Ferrer}, \citenamefont {Flambaum}, \citenamefont {Fritzsche},
  \citenamefont {Giacoppo}, \citenamefont {G{\"o}tz}, \citenamefont
  {He{\ss}berger}, \citenamefont {Huyse}, \citenamefont {Kaldor}, \citenamefont
  {Kaleja}, \citenamefont {Khuyagbaatar}, \citenamefont {Kunz}, \citenamefont
  {Laatiaoui}, \citenamefont {Lautenschl{\"a}ger}, \citenamefont {Lauth},
  \citenamefont {Mistry}, \citenamefont {Minaya~Ramirez}, \citenamefont
  {Nazarewicz}, \citenamefont {Porsev}, \citenamefont {Safronova},
  \citenamefont {Safronova}, \citenamefont {Schuetrumpf}, \citenamefont
  {Van~Duppen}, \citenamefont {Walther}, \citenamefont {Wraith},\ and\
  \citenamefont {Yakushev}}]{2018_RaederS_PhysRevLett}%
  \BibitemOpen
  \bibfield  {author} {\bibinfo {author} {\bibfnamefont {S.}~\bibnamefont
  {Raeder}}, \bibinfo {author} {\bibfnamefont {D.}~\bibnamefont {Ackermann}},
  \bibinfo {author} {\bibfnamefont {H.}~\bibnamefont {Backe}}, \bibinfo
  {author} {\bibfnamefont {R.}~\bibnamefont {Beerwerth}}, \bibinfo {author}
  {\bibfnamefont {J.~C.}\ \bibnamefont {Berengut}}, \bibinfo {author}
  {\bibfnamefont {M.}~\bibnamefont {Block}}, \bibinfo {author} {\bibfnamefont
  {A.}~\bibnamefont {Borschevsky}}, \bibinfo {author} {\bibfnamefont
  {B.}~\bibnamefont {Cheal}}, \bibinfo {author} {\bibfnamefont
  {P.}~\bibnamefont {Chhetri}}, \bibinfo {author} {\bibfnamefont {C.~E.}\
  \bibnamefont {D{\"u}llmann}}, \bibinfo {author} {\bibfnamefont {V.~A.}\
  \bibnamefont {Dzuba}}, \bibinfo {author} {\bibfnamefont {E.}~\bibnamefont
  {Eliav}}, \bibinfo {author} {\bibfnamefont {J.}~\bibnamefont {Even}},
  \bibinfo {author} {\bibfnamefont {R.}~\bibnamefont {Ferrer}}, \bibinfo
  {author} {\bibfnamefont {V.~V.}\ \bibnamefont {Flambaum}}, \bibinfo {author}
  {\bibfnamefont {S.}~\bibnamefont {Fritzsche}}, \bibinfo {author}
  {\bibfnamefont {F.}~\bibnamefont {Giacoppo}}, \bibinfo {author}
  {\bibfnamefont {S.}~\bibnamefont {G{\"o}tz}}, \bibinfo {author}
  {\bibfnamefont {F.~P.}\ \bibnamefont {He{\ss}berger}}, \bibinfo {author}
  {\bibfnamefont {M.}~\bibnamefont {Huyse}}, \bibinfo {author} {\bibfnamefont
  {U.}~\bibnamefont {Kaldor}}, \bibinfo {author} {\bibfnamefont
  {O.}~\bibnamefont {Kaleja}}, \bibinfo {author} {\bibfnamefont
  {J.}~\bibnamefont {Khuyagbaatar}}, \bibinfo {author} {\bibfnamefont
  {P.}~\bibnamefont {Kunz}}, \bibinfo {author} {\bibfnamefont {M.}~\bibnamefont
  {Laatiaoui}}, \bibinfo {author} {\bibfnamefont {F.}~\bibnamefont
  {Lautenschl{\"a}ger}}, \bibinfo {author} {\bibfnamefont {W.}~\bibnamefont
  {Lauth}}, \bibinfo {author} {\bibfnamefont {A.~K.}\ \bibnamefont {Mistry}},
  \bibinfo {author} {\bibfnamefont {E.}~\bibnamefont {Minaya~Ramirez}},
  \bibinfo {author} {\bibfnamefont {W.}~\bibnamefont {Nazarewicz}}, \bibinfo
  {author} {\bibfnamefont {S.~G.}\ \bibnamefont {Porsev}}, \bibinfo {author}
  {\bibfnamefont {M.~S.}\ \bibnamefont {Safronova}}, \bibinfo {author}
  {\bibfnamefont {U.~I.}\ \bibnamefont {Safronova}}, \bibinfo {author}
  {\bibfnamefont {B.}~\bibnamefont {Schuetrumpf}}, \bibinfo {author}
  {\bibfnamefont {P.}~\bibnamefont {Van~Duppen}}, \bibinfo {author}
  {\bibfnamefont {T.}~\bibnamefont {Walther}}, \bibinfo {author} {\bibfnamefont
  {C.}~\bibnamefont {Wraith}},\ and\ \bibinfo {author} {\bibfnamefont
  {A.}~\bibnamefont {Yakushev}},\ }\href
  {https://doi.org/10.1103/PhysRevLett.120.232503} {\bibfield  {journal}
  {\bibinfo  {journal} {Physical Review Letters}\ }\textbf {\bibinfo {volume}
  {120}},\ \bibinfo {pages} {232503} (\bibinfo {year} {2018})}\BibitemShut
  {NoStop}%
\bibitem [{\citenamefont {D{\"u}llmann}\ \emph {et~al.}(2002)\citenamefont
  {D{\"u}llmann}, \citenamefont {Br{\"u}chle}, \citenamefont {Dressler},
  \citenamefont {Eberhardt}, \citenamefont {Eichler}, \citenamefont {Eichler},
  \citenamefont {G{\"a}ggeler}, \citenamefont {Ginter}, \citenamefont {Glaus},
  \citenamefont {Gregorich}, \citenamefont {Hoffman}, \citenamefont
  {J{\"a}ger}, \citenamefont {Jost}, \citenamefont {Kirbach}, \citenamefont
  {Lee}, \citenamefont {Nitsche}, \citenamefont {Patin}, \citenamefont
  {Pershina}, \citenamefont {Piguet}, \citenamefont {Qin}, \citenamefont
  {Sch{\"a}del}, \citenamefont {Schausten}, \citenamefont {Schimpf},
  \citenamefont {Sch{\"o}tt}, \citenamefont {Soverna}, \citenamefont {Sudowe},
  \citenamefont {Th{\"o}rle}, \citenamefont {Timokhin}, \citenamefont
  {Trautmann}, \citenamefont {T{\"u}rler}, \citenamefont {Vahle}, \citenamefont
  {Wirth}, \citenamefont {Yakushev},\ and\ \citenamefont
  {Zielinski}}]{2002_DullmannC_Nature}%
  \BibitemOpen
  \bibfield  {author} {\bibinfo {author} {\bibfnamefont {C.~E.}\ \bibnamefont
  {D{\"u}llmann}}, \bibinfo {author} {\bibfnamefont {W.}~\bibnamefont
  {Br{\"u}chle}}, \bibinfo {author} {\bibfnamefont {R.}~\bibnamefont
  {Dressler}}, \bibinfo {author} {\bibfnamefont {K.}~\bibnamefont {Eberhardt}},
  \bibinfo {author} {\bibfnamefont {B.}~\bibnamefont {Eichler}}, \bibinfo
  {author} {\bibfnamefont {R.}~\bibnamefont {Eichler}}, \bibinfo {author}
  {\bibfnamefont {H.~W.}\ \bibnamefont {G{\"a}ggeler}}, \bibinfo {author}
  {\bibfnamefont {T.~N.}\ \bibnamefont {Ginter}}, \bibinfo {author}
  {\bibfnamefont {F.}~\bibnamefont {Glaus}}, \bibinfo {author} {\bibfnamefont
  {K.~E.}\ \bibnamefont {Gregorich}}, \bibinfo {author} {\bibfnamefont {D.~C.}\
  \bibnamefont {Hoffman}}, \bibinfo {author} {\bibfnamefont {E.}~\bibnamefont
  {J{\"a}ger}}, \bibinfo {author} {\bibfnamefont {D.~T.}\ \bibnamefont {Jost}},
  \bibinfo {author} {\bibfnamefont {U.~W.}\ \bibnamefont {Kirbach}}, \bibinfo
  {author} {\bibfnamefont {D.~M.}\ \bibnamefont {Lee}}, \bibinfo {author}
  {\bibfnamefont {H.}~\bibnamefont {Nitsche}}, \bibinfo {author} {\bibfnamefont
  {J.~B.}\ \bibnamefont {Patin}}, \bibinfo {author} {\bibfnamefont
  {V.}~\bibnamefont {Pershina}}, \bibinfo {author} {\bibfnamefont
  {D.}~\bibnamefont {Piguet}}, \bibinfo {author} {\bibfnamefont
  {Z.}~\bibnamefont {Qin}}, \bibinfo {author} {\bibfnamefont {M.}~\bibnamefont
  {Sch{\"a}del}}, \bibinfo {author} {\bibfnamefont {B.}~\bibnamefont
  {Schausten}}, \bibinfo {author} {\bibfnamefont {E.}~\bibnamefont {Schimpf}},
  \bibinfo {author} {\bibfnamefont {H.-J.}\ \bibnamefont {Sch{\"o}tt}},
  \bibinfo {author} {\bibfnamefont {S.}~\bibnamefont {Soverna}}, \bibinfo
  {author} {\bibfnamefont {R.}~\bibnamefont {Sudowe}}, \bibinfo {author}
  {\bibfnamefont {P.}~\bibnamefont {Th{\"o}rle}}, \bibinfo {author}
  {\bibfnamefont {S.~N.}\ \bibnamefont {Timokhin}}, \bibinfo {author}
  {\bibfnamefont {N.}~\bibnamefont {Trautmann}}, \bibinfo {author}
  {\bibfnamefont {A.}~\bibnamefont {T{\"u}rler}}, \bibinfo {author}
  {\bibfnamefont {A.}~\bibnamefont {Vahle}}, \bibinfo {author} {\bibfnamefont
  {G.}~\bibnamefont {Wirth}}, \bibinfo {author} {\bibfnamefont {A.~B.}\
  \bibnamefont {Yakushev}},\ and\ \bibinfo {author} {\bibfnamefont {P.~M.}\
  \bibnamefont {Zielinski}},\ }\href {https://doi.org/10.1038/nature00980}
  {\bibfield  {journal} {\bibinfo  {journal} {Nature}\ }\textbf {\bibinfo
  {volume} {418}},\ \bibinfo {pages} {859} (\bibinfo {year}
  {2002})}\BibitemShut {NoStop}%
\bibitem [{\citenamefont {Eichler}\ \emph {et~al.}(2007)\citenamefont
  {Eichler}, \citenamefont {Aksenov}, \citenamefont {Belozerov}, \citenamefont
  {Bozhikov}, \citenamefont {Chepigin}, \citenamefont {Dmitriev}, \citenamefont
  {Dressler}, \citenamefont {G{\"a}ggeler}, \citenamefont {Gorshkov},
  \citenamefont {Haenssler}, \citenamefont {Itkis}, \citenamefont {Laube},
  \citenamefont {Lebedev}, \citenamefont {Malyshev}, \citenamefont
  {Oganessian}, \citenamefont {Petrushkin}, \citenamefont {Piguet},
  \citenamefont {Rasmussen}, \citenamefont {Shishkin}, \citenamefont {Shutov},
  \citenamefont {Svirikhin}, \citenamefont {Tereshatov}, \citenamefont
  {Vostokin}, \citenamefont {Wegrzecki},\ and\ \citenamefont
  {Yeremin}}]{2007_EichlerR_Nature}%
  \BibitemOpen
  \bibfield  {author} {\bibinfo {author} {\bibfnamefont {R.}~\bibnamefont
  {Eichler}}, \bibinfo {author} {\bibfnamefont {N.~V.}\ \bibnamefont
  {Aksenov}}, \bibinfo {author} {\bibfnamefont {A.~V.}\ \bibnamefont
  {Belozerov}}, \bibinfo {author} {\bibfnamefont {G.~A.}\ \bibnamefont
  {Bozhikov}}, \bibinfo {author} {\bibfnamefont {V.~I.}\ \bibnamefont
  {Chepigin}}, \bibinfo {author} {\bibfnamefont {S.~N.}\ \bibnamefont
  {Dmitriev}}, \bibinfo {author} {\bibfnamefont {R.}~\bibnamefont {Dressler}},
  \bibinfo {author} {\bibfnamefont {H.~W.}\ \bibnamefont {G{\"a}ggeler}},
  \bibinfo {author} {\bibfnamefont {V.~A.}\ \bibnamefont {Gorshkov}}, \bibinfo
  {author} {\bibfnamefont {F.}~\bibnamefont {Haenssler}}, \bibinfo {author}
  {\bibfnamefont {M.~G.}\ \bibnamefont {Itkis}}, \bibinfo {author}
  {\bibfnamefont {A.}~\bibnamefont {Laube}}, \bibinfo {author} {\bibfnamefont
  {V.~Y.}\ \bibnamefont {Lebedev}}, \bibinfo {author} {\bibfnamefont {O.~N.}\
  \bibnamefont {Malyshev}}, \bibinfo {author} {\bibfnamefont {Y.~T.}\
  \bibnamefont {Oganessian}}, \bibinfo {author} {\bibfnamefont {O.~V.}\
  \bibnamefont {Petrushkin}}, \bibinfo {author} {\bibfnamefont
  {D.}~\bibnamefont {Piguet}}, \bibinfo {author} {\bibfnamefont
  {P.}~\bibnamefont {Rasmussen}}, \bibinfo {author} {\bibfnamefont {S.~V.}\
  \bibnamefont {Shishkin}}, \bibinfo {author} {\bibfnamefont {A.~V.}\
  \bibnamefont {Shutov}}, \bibinfo {author} {\bibfnamefont {A.~I.}\
  \bibnamefont {Svirikhin}}, \bibinfo {author} {\bibfnamefont {E.~E.}\
  \bibnamefont {Tereshatov}}, \bibinfo {author} {\bibfnamefont {G.~K.}\
  \bibnamefont {Vostokin}}, \bibinfo {author} {\bibfnamefont {M.}~\bibnamefont
  {Wegrzecki}},\ and\ \bibinfo {author} {\bibfnamefont {A.~V.}\ \bibnamefont
  {Yeremin}},\ }\href {https://doi.org/10.1038/nature05761} {\bibfield
  {journal} {\bibinfo  {journal} {Nature}\ }\textbf {\bibinfo {volume} {447}},\
  \bibinfo {pages} {72} (\bibinfo {year} {2007})}\BibitemShut {NoStop}%
\bibitem [{\citenamefont {Eichler}\ \emph {et~al.}(2008)\citenamefont
  {Eichler}, \citenamefont {Aksenov}, \citenamefont {Belozerov}, \citenamefont
  {Bozhikov}, \citenamefont {Chepigin}, \citenamefont {Dmitriev}, \citenamefont
  {Dressler}, \citenamefont {G{\"a}ggeler}, \citenamefont {Gorshkov},
  \citenamefont {Itkis}, \citenamefont {Haenssler}, \citenamefont {Laube},
  \citenamefont {Lebedev}, \citenamefont {Malyshev}, \citenamefont
  {Oganessian}, \citenamefont {Petrushkin}, \citenamefont {Piguet},
  \citenamefont {Popeko}, \citenamefont {Rasmussen}, \citenamefont {Shishkin},
  \citenamefont {Serov}, \citenamefont {Shutov}, \citenamefont {Svirikhin},
  \citenamefont {Tereshatov}, \citenamefont {Vostokin}, \citenamefont
  {Wegrzecki},\ and\ \citenamefont {Yeremin}}]{2008_EichlerR_AngewChemIntEd}%
  \BibitemOpen
  \bibfield  {author} {\bibinfo {author} {\bibfnamefont {R.}~\bibnamefont
  {Eichler}}, \bibinfo {author} {\bibfnamefont {N.~V.}\ \bibnamefont
  {Aksenov}}, \bibinfo {author} {\bibfnamefont {A.~V.}\ \bibnamefont
  {Belozerov}}, \bibinfo {author} {\bibfnamefont {G.~A.}\ \bibnamefont
  {Bozhikov}}, \bibinfo {author} {\bibfnamefont {V.~I.}\ \bibnamefont
  {Chepigin}}, \bibinfo {author} {\bibfnamefont {S.~N.}\ \bibnamefont
  {Dmitriev}}, \bibinfo {author} {\bibfnamefont {R.}~\bibnamefont {Dressler}},
  \bibinfo {author} {\bibfnamefont {H.~W.}\ \bibnamefont {G{\"a}ggeler}},
  \bibinfo {author} {\bibfnamefont {A.~V.}\ \bibnamefont {Gorshkov}}, \bibinfo
  {author} {\bibfnamefont {M.~G.}\ \bibnamefont {Itkis}}, \bibinfo {author}
  {\bibfnamefont {F.}~\bibnamefont {Haenssler}}, \bibinfo {author}
  {\bibfnamefont {A.}~\bibnamefont {Laube}}, \bibinfo {author} {\bibfnamefont
  {V.~Y.}\ \bibnamefont {Lebedev}}, \bibinfo {author} {\bibfnamefont {O.~N.}\
  \bibnamefont {Malyshev}}, \bibinfo {author} {\bibfnamefont {Y.~T.}\
  \bibnamefont {Oganessian}}, \bibinfo {author} {\bibfnamefont {O.~V.}\
  \bibnamefont {Petrushkin}}, \bibinfo {author} {\bibfnamefont
  {D.}~\bibnamefont {Piguet}}, \bibinfo {author} {\bibfnamefont {A.~G.}\
  \bibnamefont {Popeko}}, \bibinfo {author} {\bibfnamefont {P.}~\bibnamefont
  {Rasmussen}}, \bibinfo {author} {\bibfnamefont {S.~V.}\ \bibnamefont
  {Shishkin}}, \bibinfo {author} {\bibfnamefont {A.~A.}\ \bibnamefont {Serov}},
  \bibinfo {author} {\bibfnamefont {A.~V.}\ \bibnamefont {Shutov}}, \bibinfo
  {author} {\bibfnamefont {A.~I.}\ \bibnamefont {Svirikhin}}, \bibinfo {author}
  {\bibfnamefont {E.~E.}\ \bibnamefont {Tereshatov}}, \bibinfo {author}
  {\bibfnamefont {G.~K.}\ \bibnamefont {Vostokin}}, \bibinfo {author}
  {\bibfnamefont {M.}~\bibnamefont {Wegrzecki}},\ and\ \bibinfo {author}
  {\bibfnamefont {A.~V.}\ \bibnamefont {Yeremin}},\ }\href
  {https://doi.org/10.1002/anie.200705019} {\bibfield  {journal} {\bibinfo
  {journal} {Angewandte Chemie International Edition}\ }\textbf {\bibinfo
  {volume} {47}},\ \bibinfo {pages} {3262} (\bibinfo {year}
  {2008})}\BibitemShut {NoStop}%
\bibitem [{\citenamefont {Eichler}\ \emph {et~al.}(2010)\citenamefont
  {Eichler}, \citenamefont {Aksenov}, \citenamefont {Albin}, \citenamefont
  {Belozerov}, \citenamefont {Bozhikov}, \citenamefont {Chepigin},
  \citenamefont {Dmitriev}, \citenamefont {Dressler}, \citenamefont
  {G{\"a}ggeler}, \citenamefont {Gorshkov}, \citenamefont {Henderson},\ and\
  \citenamefont {Al}}]{2010_EichlerR_RadChimActa}%
  \BibitemOpen
  \bibfield  {author} {\bibinfo {author} {\bibfnamefont {R.}~\bibnamefont
  {Eichler}}, \bibinfo {author} {\bibfnamefont {N.~V.}\ \bibnamefont
  {Aksenov}}, \bibinfo {author} {\bibfnamefont {Y.~V.}\ \bibnamefont {Albin}},
  \bibinfo {author} {\bibfnamefont {A.~V.}\ \bibnamefont {Belozerov}}, \bibinfo
  {author} {\bibfnamefont {G.~A.}\ \bibnamefont {Bozhikov}}, \bibinfo {author}
  {\bibfnamefont {V.~I.}\ \bibnamefont {Chepigin}}, \bibinfo {author}
  {\bibfnamefont {S.~N.}\ \bibnamefont {Dmitriev}}, \bibinfo {author}
  {\bibfnamefont {R.}~\bibnamefont {Dressler}}, \bibinfo {author}
  {\bibfnamefont {H.~W.}\ \bibnamefont {G{\"a}ggeler}}, \bibinfo {author}
  {\bibfnamefont {V.~A.}\ \bibnamefont {Gorshkov}}, \bibinfo {author}
  {\bibfnamefont {G.~S.}\ \bibnamefont {Henderson}},\ and\ \bibinfo {author}
  {\bibfnamefont {E.}~\bibnamefont {Al}},\ }\href@noop {} {\bibfield  {journal}
  {\bibinfo  {journal} {Radiochimica Acta}\ }\textbf {\bibinfo {volume} {98}},\
  \bibinfo {pages} {133} (\bibinfo {year} {2010})}\BibitemShut {NoStop}%
\bibitem [{\citenamefont {Dmitriev}\ \emph {et~al.}(2014)\citenamefont
  {Dmitriev}, \citenamefont {Aksenov}, \citenamefont {Albin}, \citenamefont
  {Bozhikov}, \citenamefont {Chelnokov}, \citenamefont {Chepygin},
  \citenamefont {Eichler}, \citenamefont {Isaev}, \citenamefont {Katrasev},
  \citenamefont {Lebedev}, \citenamefont {Malyshev}, \citenamefont
  {Petrushkin}, \citenamefont {Porobanuk}, \citenamefont {Ryabinin},
  \citenamefont {Sabel'nikov}, \citenamefont {Sokol}, \citenamefont
  {Svirikhin}, \citenamefont {Starodub}, \citenamefont {Usoltsev},
  \citenamefont {Vostokin},\ and\ \citenamefont
  {Yeremin}}]{2014_DmitrievS_MendComm}%
  \BibitemOpen
  \bibfield  {author} {\bibinfo {author} {\bibfnamefont {S.~N.}\ \bibnamefont
  {Dmitriev}}, \bibinfo {author} {\bibfnamefont {N.~V.}\ \bibnamefont
  {Aksenov}}, \bibinfo {author} {\bibfnamefont {Y.~V.}\ \bibnamefont {Albin}},
  \bibinfo {author} {\bibfnamefont {G.~A.}\ \bibnamefont {Bozhikov}}, \bibinfo
  {author} {\bibfnamefont {M.~L.}\ \bibnamefont {Chelnokov}}, \bibinfo {author}
  {\bibfnamefont {V.~I.}\ \bibnamefont {Chepygin}}, \bibinfo {author}
  {\bibfnamefont {R.}~\bibnamefont {Eichler}}, \bibinfo {author} {\bibfnamefont
  {A.~V.}\ \bibnamefont {Isaev}}, \bibinfo {author} {\bibfnamefont {D.~E.}\
  \bibnamefont {Katrasev}}, \bibinfo {author} {\bibfnamefont {V.~Y.}\
  \bibnamefont {Lebedev}}, \bibinfo {author} {\bibfnamefont {O.~N.}\
  \bibnamefont {Malyshev}}, \bibinfo {author} {\bibfnamefont {O.~V.}\
  \bibnamefont {Petrushkin}}, \bibinfo {author} {\bibfnamefont {L.~S.}\
  \bibnamefont {Porobanuk}}, \bibinfo {author} {\bibfnamefont {M.~A.}\
  \bibnamefont {Ryabinin}}, \bibinfo {author} {\bibfnamefont {A.~V.}\
  \bibnamefont {Sabel'nikov}}, \bibinfo {author} {\bibfnamefont {E.~A.}\
  \bibnamefont {Sokol}}, \bibinfo {author} {\bibfnamefont {A.~V.}\ \bibnamefont
  {Svirikhin}}, \bibinfo {author} {\bibfnamefont {G.~Y.}\ \bibnamefont
  {Starodub}}, \bibinfo {author} {\bibfnamefont {I.}~\bibnamefont {Usoltsev}},
  \bibinfo {author} {\bibfnamefont {G.~K.}\ \bibnamefont {Vostokin}},\ and\
  \bibinfo {author} {\bibfnamefont {A.~V.}\ \bibnamefont {Yeremin}},\ }\href
  {https://doi.org/10.1016/j.mencom.2014.09.001} {\bibfield  {journal}
  {\bibinfo  {journal} {Mendeleev Communications}\ }\textbf {\bibinfo {volume}
  {24}},\ \bibinfo {pages} {253} (\bibinfo {year} {2014})}\BibitemShut
  {NoStop}%
\bibitem [{\citenamefont {Aksenov}\ \emph {et~al.}(2017)\citenamefont
  {Aksenov}, \citenamefont {Steinegger}, \citenamefont {Abdullin},
  \citenamefont {Albin}, \citenamefont {Bozhikov}, \citenamefont {Chepigin},
  \citenamefont {Eichler}, \citenamefont {Lebedev}, \citenamefont {Madumarov},
  \citenamefont {Malyshev}, \citenamefont {Petrushkin}, \citenamefont
  {Polyakov}, \citenamefont {Popov}, \citenamefont {Sabel'nikov}, \citenamefont
  {Sagaidak}, \citenamefont {Shirokovsky}, \citenamefont {Shumeiko},
  \citenamefont {Starodub}, \citenamefont {Tsyganov}, \citenamefont {Utyonkov},
  \citenamefont {Voinov}, \citenamefont {Vostokin}, \citenamefont {Yeremin},\
  and\ \citenamefont {Dmitriev}}]{2017_AksenovN_EurPhysJA}%
  \BibitemOpen
  \bibfield  {author} {\bibinfo {author} {\bibfnamefont {N.~V.}\ \bibnamefont
  {Aksenov}}, \bibinfo {author} {\bibfnamefont {P.}~\bibnamefont {Steinegger}},
  \bibinfo {author} {\bibfnamefont {F.~S.}\ \bibnamefont {Abdullin}}, \bibinfo
  {author} {\bibfnamefont {Y.~V.}\ \bibnamefont {Albin}}, \bibinfo {author}
  {\bibfnamefont {G.~A.}\ \bibnamefont {Bozhikov}}, \bibinfo {author}
  {\bibfnamefont {V.~I.}\ \bibnamefont {Chepigin}}, \bibinfo {author}
  {\bibfnamefont {R.}~\bibnamefont {Eichler}}, \bibinfo {author} {\bibfnamefont
  {V.~Y.}\ \bibnamefont {Lebedev}}, \bibinfo {author} {\bibfnamefont {A.~S.}\
  \bibnamefont {Madumarov}}, \bibinfo {author} {\bibfnamefont {O.~N.}\
  \bibnamefont {Malyshev}}, \bibinfo {author} {\bibfnamefont {O.~V.}\
  \bibnamefont {Petrushkin}}, \bibinfo {author} {\bibfnamefont {A.~N.}\
  \bibnamefont {Polyakov}}, \bibinfo {author} {\bibfnamefont {Y.~A.}\
  \bibnamefont {Popov}}, \bibinfo {author} {\bibfnamefont {A.~V.}\ \bibnamefont
  {Sabel'nikov}}, \bibinfo {author} {\bibfnamefont {R.~N.}\ \bibnamefont
  {Sagaidak}}, \bibinfo {author} {\bibfnamefont {I.~V.}\ \bibnamefont
  {Shirokovsky}}, \bibinfo {author} {\bibfnamefont {M.~V.}\ \bibnamefont
  {Shumeiko}}, \bibinfo {author} {\bibfnamefont {G.~Y.}\ \bibnamefont
  {Starodub}}, \bibinfo {author} {\bibfnamefont {Y.~S.}\ \bibnamefont
  {Tsyganov}}, \bibinfo {author} {\bibfnamefont {V.~K.}\ \bibnamefont
  {Utyonkov}}, \bibinfo {author} {\bibfnamefont {A.~A.}\ \bibnamefont
  {Voinov}}, \bibinfo {author} {\bibfnamefont {G.~K.}\ \bibnamefont
  {Vostokin}}, \bibinfo {author} {\bibfnamefont {A.~V.}\ \bibnamefont
  {Yeremin}},\ and\ \bibinfo {author} {\bibfnamefont {S.~N.}\ \bibnamefont
  {Dmitriev}},\ }\href {https://doi.org/10.1140/epja/i2017-12348-8} {\bibfield
  {journal} {\bibinfo  {journal} {The European Physical Journal A}\ }\textbf
  {\bibinfo {volume} {53}},\ \bibinfo {pages} {158} (\bibinfo {year}
  {2017})}\BibitemShut {NoStop}%
\bibitem [{\citenamefont {Fricke}\ \emph {et~al.}(1971)\citenamefont {Fricke},
  \citenamefont {Greiner},\ and\ \citenamefont
  {Waber}}]{1971_FrickeB_ThChimAct21}%
  \BibitemOpen
  \bibfield  {author} {\bibinfo {author} {\bibfnamefont {B.}~\bibnamefont
  {Fricke}}, \bibinfo {author} {\bibfnamefont {W.}~\bibnamefont {Greiner}},\
  and\ \bibinfo {author} {\bibfnamefont {J.~T.}\ \bibnamefont {Waber}},\ }\href
  {https://doi.org/10.1007/BF01172015} {\bibfield  {journal} {\bibinfo
  {journal} {Theoretica chimica acta}\ }\textbf {\bibinfo {volume} {21}},\
  \bibinfo {pages} {235} (\bibinfo {year} {1971})}\BibitemShut {NoStop}%
\bibitem [{\citenamefont {Seaborg}(1996)}]{1996_SeaborgG_JChemSoc}%
  \BibitemOpen
  \bibfield  {author} {\bibinfo {author} {\bibfnamefont {G.~T.}\ \bibnamefont
  {Seaborg}},\ }\href {https://doi.org/10.1039/DT9960003899} {\bibfield
  {journal} {\bibinfo  {journal} {Journal of the Chemical Society, Dalton
  Transactions}\ ,\ \bibinfo {pages} {3899}} (\bibinfo {year}
  {1996})}\BibitemShut {NoStop}%
\bibitem [{\citenamefont {Nefedov}\ \emph {et~al.}(2006)\citenamefont
  {Nefedov}, \citenamefont {Trzhaskovskaya},\ and\ \citenamefont
  {Yarzhemskii}}]{2006_NefedovV_DoklPhysChem}%
  \BibitemOpen
  \bibfield  {author} {\bibinfo {author} {\bibfnamefont {V.~I.}\ \bibnamefont
  {Nefedov}}, \bibinfo {author} {\bibfnamefont {M.~B.}\ \bibnamefont
  {Trzhaskovskaya}},\ and\ \bibinfo {author} {\bibfnamefont {V.~G.}\
  \bibnamefont {Yarzhemskii}},\ }\href
  {https://doi.org/10.1134/S0012501606060029} {\bibfield  {journal} {\bibinfo
  {journal} {Doklady Physical Chemistry}\ }\textbf {\bibinfo {volume} {408}},\
  \bibinfo {pages} {149} (\bibinfo {year} {2006})}\BibitemShut {NoStop}%
\bibitem [{\citenamefont {Pyykk{\"o}}(2010)}]{2010_PykkoP_PhysChemChemPhys13}%
  \BibitemOpen
  \bibfield  {author} {\bibinfo {author} {\bibfnamefont {P.}~\bibnamefont
  {Pyykk{\"o}}},\ }\href {https://doi.org/10.1039/C0CP01575J} {\bibfield
  {journal} {\bibinfo  {journal} {Physical Chemistry Chemical Physics}\
  }\textbf {\bibinfo {volume} {13}},\ \bibinfo {pages} {161} (\bibinfo {year}
  {2010})}\BibitemShut {NoStop}%
\bibitem [{\citenamefont {Jerabek}\ \emph {et~al.}(2018)\citenamefont
  {Jerabek}, \citenamefont {Schuetrumpf}, \citenamefont {Schwerdtfeger},\ and\
  \citenamefont {Nazarewicz}}]{2018_JerabekP_PhysRevLett}%
  \BibitemOpen
  \bibfield  {author} {\bibinfo {author} {\bibfnamefont {P.}~\bibnamefont
  {Jerabek}}, \bibinfo {author} {\bibfnamefont {B.}~\bibnamefont
  {Schuetrumpf}}, \bibinfo {author} {\bibfnamefont {P.}~\bibnamefont
  {Schwerdtfeger}},\ and\ \bibinfo {author} {\bibfnamefont {W.}~\bibnamefont
  {Nazarewicz}},\ }\href {https://doi.org/10.1103/PhysRevLett.120.053001}
  {\bibfield  {journal} {\bibinfo  {journal} {Physical Review Letters}\
  }\textbf {\bibinfo {volume} {120}},\ \bibinfo {pages} {053001} (\bibinfo
  {year} {2018})}\BibitemShut {NoStop}%
\bibitem [{\citenamefont {Kaygorodov}\ \emph {et~al.}(2019)\citenamefont
  {Kaygorodov}, \citenamefont {Kozhedub}, \citenamefont {I.~Tupitsyn},\ and\
  \citenamefont {M.~Shabaev}}]{2019_KaygorodovM_POS}%
  \BibitemOpen
  \bibfield  {author} {\bibinfo {author} {\bibfnamefont {M.~Y.}\ \bibnamefont
  {Kaygorodov}}, \bibinfo {author} {\bibfnamefont {Y.~S.}\ \bibnamefont
  {Kozhedub}}, \bibinfo {author} {\bibfnamefont {I.}~\bibnamefont
  {I.~Tupitsyn}},\ and\ \bibinfo {author} {\bibfnamefont {V.}~\bibnamefont
  {M.~Shabaev}},\ }\href@noop {} {\bibfield  {journal} {\bibinfo  {journal}
  {Proceedings of Science}\ }\textbf {\bibinfo {volume} {353}},\ \bibinfo
  {pages} {036} (\bibinfo {year} {2019})}\BibitemShut {NoStop}%
\bibitem [{\citenamefont {Mann}\ and\ \citenamefont
  {Waber}(1970)}]{1970_MannJ_JChemPhys}%
  \BibitemOpen
  \bibfield  {author} {\bibinfo {author} {\bibfnamefont {J.~B.}\ \bibnamefont
  {Mann}}\ and\ \bibinfo {author} {\bibfnamefont {J.~T.}\ \bibnamefont
  {Waber}},\ }\href {https://doi.org/10.1063/1.1674338} {\bibfield  {journal}
  {\bibinfo  {journal} {The Journal of Chemical Physics}\ }\textbf {\bibinfo
  {volume} {53}},\ \bibinfo {pages} {2397} (\bibinfo {year}
  {1970})}\BibitemShut {NoStop}%
\bibitem [{\citenamefont {Landau}\ \emph {et~al.}(2001)\citenamefont {Landau},
  \citenamefont {Eliav}, \citenamefont {Ishikawa},\ and\ \citenamefont
  {Kaldor}}]{2001_LandauA_JChemPhys}%
  \BibitemOpen
  \bibfield  {author} {\bibinfo {author} {\bibfnamefont {A.}~\bibnamefont
  {Landau}}, \bibinfo {author} {\bibfnamefont {E.}~\bibnamefont {Eliav}},
  \bibinfo {author} {\bibfnamefont {Y.}~\bibnamefont {Ishikawa}},\ and\
  \bibinfo {author} {\bibfnamefont {U.}~\bibnamefont {Kaldor}},\ }\href
  {https://doi.org/10.1063/1.1386413} {\bibfield  {journal} {\bibinfo
  {journal} {The Journal of Chemical Physics}\ }\textbf {\bibinfo {volume}
  {115}},\ \bibinfo {pages} {2389} (\bibinfo {year} {2001})}\BibitemShut
  {NoStop}%
\bibitem [{\citenamefont {Eliav}\ \emph {et~al.}(2002)\citenamefont {Eliav},
  \citenamefont {Landau}, \citenamefont {Ishikawa},\ and\ \citenamefont
  {Kaldor}}]{2002_EliavE_JPhysB}%
  \BibitemOpen
  \bibfield  {author} {\bibinfo {author} {\bibfnamefont {E.}~\bibnamefont
  {Eliav}}, \bibinfo {author} {\bibfnamefont {A.}~\bibnamefont {Landau}},
  \bibinfo {author} {\bibfnamefont {Y.}~\bibnamefont {Ishikawa}},\ and\
  \bibinfo {author} {\bibfnamefont {U.}~\bibnamefont {Kaldor}},\ }\href
  {https://doi.org/10.1088/0953-4075/35/7/307} {\bibfield  {journal} {\bibinfo
  {journal} {Journal of Physics B: Atomic, Molecular and Optical Physics}\
  }\textbf {\bibinfo {volume} {35}},\ \bibinfo {pages} {1693} (\bibinfo {year}
  {2002})}\BibitemShut {NoStop}%
\bibitem [{\citenamefont {Mosyagin}\ \emph {et~al.}(2006)\citenamefont
  {Mosyagin}, \citenamefont {Isaev},\ and\ \citenamefont
  {Titov}}]{Mosyagin:06b}%
  \BibitemOpen
  \bibfield  {author} {\bibinfo {author} {\bibfnamefont {N.~S.}\ \bibnamefont
  {Mosyagin}}, \bibinfo {author} {\bibfnamefont {T.~A.}\ \bibnamefont
  {Isaev}},\ and\ \bibinfo {author} {\bibfnamefont {A.~V.}\ \bibnamefont
  {Titov}},\ }\href {https://doi.org/10.1063/1.2206189} {\bibfield  {journal}
  {\bibinfo  {journal} {J.\ Chem.\ Phys.}\ }\textbf {\bibinfo {volume} {124}},\
  \bibinfo {pages} {224302} (\bibinfo {year} {2006})}\BibitemShut {NoStop}%
\bibitem [{\citenamefont {Zaitsevskii}\ \emph {et~al.}(2006)\citenamefont
  {Zaitsevskii}, \citenamefont {Rykova}, \citenamefont {Mosyagin},\ and\
  \citenamefont {Titov}}]{Zaitsevskii:06a}%
  \BibitemOpen
  \bibfield  {author} {\bibinfo {author} {\bibfnamefont {A.~V.}\ \bibnamefont
  {Zaitsevskii}}, \bibinfo {author} {\bibfnamefont {E.~A.}\ \bibnamefont
  {Rykova}}, \bibinfo {author} {\bibfnamefont {N.~S.}\ \bibnamefont
  {Mosyagin}},\ and\ \bibinfo {author} {\bibfnamefont {A.~V.}\ \bibnamefont
  {Titov}},\ }\href {https://doi.org/10.2478/s11534-006-0029-7} {\ \textbf
  {\bibinfo {volume} {4}},\ \bibinfo {pages} {448} (\bibinfo {year}
  {2006})}\BibitemShut {NoStop}%
\bibitem [{\citenamefont {Liu}\ \emph {et~al.}(2007)\citenamefont {Liu},
  \citenamefont {Hutton},\ and\ \citenamefont {Zou}}]{2007_LiuY_PhysRevA}%
  \BibitemOpen
  \bibfield  {author} {\bibinfo {author} {\bibfnamefont {Y.}~\bibnamefont
  {Liu}}, \bibinfo {author} {\bibfnamefont {R.}~\bibnamefont {Hutton}},\ and\
  \bibinfo {author} {\bibfnamefont {Y.}~\bibnamefont {Zou}},\ }\href
  {https://doi.org/10.1103/PhysRevA.76.062503} {\bibfield  {journal} {\bibinfo
  {journal} {Physical Review A}\ }\textbf {\bibinfo {volume} {76}},\ \bibinfo
  {pages} {062503} (\bibinfo {year} {2007})}\BibitemShut {NoStop}%
\bibitem [{\citenamefont {Dinh}\ \emph {et~al.}(2008)\citenamefont {Dinh},
  \citenamefont {Dzuba}, \citenamefont {Flambaum},\ and\ \citenamefont
  {Ginges}}]{2008_DinhT_PhysRevA78}%
  \BibitemOpen
  \bibfield  {author} {\bibinfo {author} {\bibfnamefont {T.~H.}\ \bibnamefont
  {Dinh}}, \bibinfo {author} {\bibfnamefont {V.~A.}\ \bibnamefont {Dzuba}},
  \bibinfo {author} {\bibfnamefont {V.~V.}\ \bibnamefont {Flambaum}},\ and\
  \bibinfo {author} {\bibfnamefont {J.~S.~M.}\ \bibnamefont {Ginges}},\ }\href
  {https://doi.org/10.1103/PhysRevA.78.022507} {\bibfield  {journal} {\bibinfo
  {journal} {Physical Review A}\ }\textbf {\bibinfo {volume} {78}},\ \bibinfo
  {pages} {022507} (\bibinfo {year} {2008})}\BibitemShut {NoStop}%
\bibitem [{\citenamefont {Pershina}\ \emph {et~al.}(2008)\citenamefont
  {Pershina}, \citenamefont {Borschevsky}, \citenamefont {Eliav},\ and\
  \citenamefont {Kaldor}}]{2008_PershinaV_JChemPhys129}%
  \BibitemOpen
  \bibfield  {author} {\bibinfo {author} {\bibfnamefont {V.}~\bibnamefont
  {Pershina}}, \bibinfo {author} {\bibfnamefont {A.}~\bibnamefont
  {Borschevsky}}, \bibinfo {author} {\bibfnamefont {E.}~\bibnamefont {Eliav}},\
  and\ \bibinfo {author} {\bibfnamefont {U.}~\bibnamefont {Kaldor}},\ }\href
  {https://doi.org/10.1063/1.2988318} {\bibfield  {journal} {\bibinfo
  {journal} {The Journal of Chemical Physics}\ }\textbf {\bibinfo {volume}
  {129}},\ \bibinfo {pages} {144106} (\bibinfo {year} {2008})}\BibitemShut
  {NoStop}%
\bibitem [{\citenamefont {Borschevsky}\ \emph {et~al.}(2009)\citenamefont
  {Borschevsky}, \citenamefont {Pershina}, \citenamefont {Eliav},\ and\
  \citenamefont {Kaldor}}]{2009_BorschevskyA_ChemPhysLett}%
  \BibitemOpen
  \bibfield  {author} {\bibinfo {author} {\bibfnamefont {A.}~\bibnamefont
  {Borschevsky}}, \bibinfo {author} {\bibfnamefont {V.}~\bibnamefont
  {Pershina}}, \bibinfo {author} {\bibfnamefont {E.}~\bibnamefont {Eliav}},\
  and\ \bibinfo {author} {\bibfnamefont {U.}~\bibnamefont {Kaldor}},\ }\href
  {https://doi.org/10.1016/j.cplett.2009.08.059} {\bibfield  {journal}
  {\bibinfo  {journal} {Chemical Physics Letters}\ }\textbf {\bibinfo {volume}
  {480}},\ \bibinfo {pages} {49} (\bibinfo {year} {2009})}\BibitemShut
  {NoStop}%
\bibitem [{\citenamefont {Goidenko}(2009)}]{2009_GoidenkoI_EurPhysJD}%
  \BibitemOpen
  \bibfield  {author} {\bibinfo {author} {\bibfnamefont {I.~A.}\ \bibnamefont
  {Goidenko}},\ }\href {https://doi.org/10.1140/epjd/e2009-00216-4} {\bibfield
  {journal} {\bibinfo  {journal} {The European Physical Journal D}\ }\textbf
  {\bibinfo {volume} {55}},\ \bibinfo {pages} {35} (\bibinfo {year}
  {2009})}\BibitemShut {NoStop}%
\bibitem [{\citenamefont {Zaitsevskii}\ \emph {et~al.}(2009)\citenamefont
  {Zaitsevskii}, \citenamefont {van W{\"u}llen},\ and\ \citenamefont
  {Titov}}]{2009_ZaitsevskiiA_RussChemRev}%
  \BibitemOpen
  \bibfield  {author} {\bibinfo {author} {\bibfnamefont {A.~V.}\ \bibnamefont
  {Zaitsevskii}}, \bibinfo {author} {\bibfnamefont {C.}~\bibnamefont {van
  W{\"u}llen}},\ and\ \bibinfo {author} {\bibfnamefont {A.~V.}\ \bibnamefont
  {Titov}},\ }\href {https://doi.org/10.1070/RC2009v078n12ABEH004075}
  {\bibfield  {journal} {\bibinfo  {journal} {Russian Chemical Reviews}\
  }\textbf {\bibinfo {volume} {78}},\ \bibinfo {pages} {1173} (\bibinfo {year}
  {2009})}\BibitemShut {NoStop}%
\bibitem [{\citenamefont {Zaitsevskii}\ and\ \citenamefont
  {Titov}(2013)}]{Zaitsevskii:13b}%
  \BibitemOpen
  \bibfield  {author} {\bibinfo {author} {\bibfnamefont {A.}~\bibnamefont
  {Zaitsevskii}}\ and\ \bibinfo {author} {\bibfnamefont {A.~V.}\ \bibnamefont
  {Titov}},\ }\href {https://doi.org/10.1002/qua.24429} {\ \textbf {\bibinfo
  {volume} {113}},\ \bibinfo {pages} {1772} (\bibinfo {year}
  {2013})}\BibitemShut {NoStop}%
\bibitem [{\citenamefont {Demidov}\ and\ \citenamefont
  {Zaitsevskii}(2014)}]{2014_DemidovY_RussChemBull}%
  \BibitemOpen
  \bibfield  {author} {\bibinfo {author} {\bibfnamefont {Y.~A.}\ \bibnamefont
  {Demidov}}\ and\ \bibinfo {author} {\bibfnamefont {A.~V.}\ \bibnamefont
  {Zaitsevskii}},\ }\href@noop {} {\bibfield  {journal} {\bibinfo  {journal}
  {Russian Chemical Bulletin}\ }\textbf {\bibinfo {volume} {63}},\ \bibinfo
  {pages} {1647} (\bibinfo {year} {2014})}\BibitemShut {NoStop}%
\bibitem [{\citenamefont {Hangele}\ and\ \citenamefont
  {Dolg}(2014)}]{2014_HangeleT_ChemPhysLett}%
  \BibitemOpen
  \bibfield  {author} {\bibinfo {author} {\bibfnamefont {T.}~\bibnamefont
  {Hangele}}\ and\ \bibinfo {author} {\bibfnamefont {M.}~\bibnamefont {Dolg}},\
  }\href {https://doi.org/10.1016/j.cplett.2014.10.048} {\bibfield  {journal}
  {\bibinfo  {journal} {Chemical Physics Letters}\ }\textbf {\bibinfo {volume}
  {616-617}},\ \bibinfo {pages} {222} (\bibinfo {year} {2014})}\BibitemShut
  {NoStop}%
\bibitem [{\citenamefont {Dzuba}(2016)}]{2016_DzubaV_PhysRevA}%
  \BibitemOpen
  \bibfield  {author} {\bibinfo {author} {\bibfnamefont {V.~A.}\ \bibnamefont
  {Dzuba}},\ }\href {https://doi.org/10.1103/PhysRevA.93.032519} {\bibfield
  {journal} {\bibinfo  {journal} {Physical Review A}\ }\textbf {\bibinfo
  {volume} {93}},\ \bibinfo {pages} {032519} (\bibinfo {year}
  {2016})}\BibitemShut {NoStop}%
\bibitem [{\citenamefont {Ginges}\ and\ \citenamefont
  {Berengut}(2016{\natexlab{a}})}]{2016_GingesJ_PhysRevA}%
  \BibitemOpen
  \bibfield  {author} {\bibinfo {author} {\bibfnamefont {J.~S.~M.}\
  \bibnamefont {Ginges}}\ and\ \bibinfo {author} {\bibfnamefont {J.~C.}\
  \bibnamefont {Berengut}},\ }\href
  {https://doi.org/10.1103/PhysRevA.93.052509} {\bibfield  {journal} {\bibinfo
  {journal} {Physical Review A}\ }\textbf {\bibinfo {volume} {93}},\ \bibinfo
  {pages} {052509} (\bibinfo {year} {2016}{\natexlab{a}})}\BibitemShut
  {NoStop}%
\bibitem [{\citenamefont {Ginges}\ and\ \citenamefont
  {Berengut}(2016{\natexlab{b}})}]{2016_GingesJ_JPhysBAtMolOptPhys}%
  \BibitemOpen
  \bibfield  {author} {\bibinfo {author} {\bibfnamefont {J.~S.~M.}\
  \bibnamefont {Ginges}}\ and\ \bibinfo {author} {\bibfnamefont {J.~C.}\
  \bibnamefont {Berengut}},\ }\href
  {https://doi.org/10.1088/0953-4075/49/9/095001} {\bibfield  {journal}
  {\bibinfo  {journal} {Journal of Physics B: Atomic, Molecular and Optical
  Physics}\ }\textbf {\bibinfo {volume} {49}},\ \bibinfo {pages} {095001}
  (\bibinfo {year} {2016}{\natexlab{b}})}\BibitemShut {NoStop}%
\bibitem [{\citenamefont {Jerabek}\ \emph {et~al.}(2019)\citenamefont
  {Jerabek}, \citenamefont {Smits}, \citenamefont {Mewes}, \citenamefont
  {Peterson},\ and\ \citenamefont {Schwerdtfeger}}]{2019_JerabekP_JPhysChemA}%
  \BibitemOpen
  \bibfield  {author} {\bibinfo {author} {\bibfnamefont {P.}~\bibnamefont
  {Jerabek}}, \bibinfo {author} {\bibfnamefont {O.~R.}\ \bibnamefont {Smits}},
  \bibinfo {author} {\bibfnamefont {J.-M.}\ \bibnamefont {Mewes}}, \bibinfo
  {author} {\bibfnamefont {K.~A.}\ \bibnamefont {Peterson}},\ and\ \bibinfo
  {author} {\bibfnamefont {P.}~\bibnamefont {Schwerdtfeger}},\ }\href
  {https://doi.org/10.1021/acs.jpca.9b01947} {\bibfield  {journal} {\bibinfo
  {journal} {The Journal of Physical Chemistry A}\ }\textbf {\bibinfo {volume}
  {123}},\ \bibinfo {pages} {4201} (\bibinfo {year} {2019})}\BibitemShut
  {NoStop}%
\bibitem [{\citenamefont {Mewes}\ \emph {et~al.}(2019)\citenamefont {Mewes},
  \citenamefont {Jerabek}, \citenamefont {Smits},\ and\ \citenamefont
  {Schwerdtfeger}}]{2019_MewesJ_AngewChemIntEd}%
  \BibitemOpen
  \bibfield  {author} {\bibinfo {author} {\bibfnamefont {J.-M.}\ \bibnamefont
  {Mewes}}, \bibinfo {author} {\bibfnamefont {P.}~\bibnamefont {Jerabek}},
  \bibinfo {author} {\bibfnamefont {O.~R.}\ \bibnamefont {Smits}},\ and\
  \bibinfo {author} {\bibfnamefont {P.}~\bibnamefont {Schwerdtfeger}},\ }\href
  {https://doi.org/10.1002/anie.201908327} {\bibfield  {journal} {\bibinfo
  {journal} {Angewandte Chemie International Edition}\ }\textbf {\bibinfo
  {volume} {58}},\ \bibinfo {pages} {14260} (\bibinfo {year}
  {2019})}\BibitemShut {NoStop}%
\bibitem [{\citenamefont {Lackenby}\ \emph {et~al.}(2020)\citenamefont
  {Lackenby}, \citenamefont {Dzuba},\ and\ \citenamefont
  {Flambaum}}]{2020_LackenbyB_PhysRevA101}%
  \BibitemOpen
  \bibfield  {author} {\bibinfo {author} {\bibfnamefont {B.~G.~C.}\
  \bibnamefont {Lackenby}}, \bibinfo {author} {\bibfnamefont {V.~A.}\
  \bibnamefont {Dzuba}},\ and\ \bibinfo {author} {\bibfnamefont {V.~V.}\
  \bibnamefont {Flambaum}},\ }\href
  {https://doi.org/10.1103/PhysRevA.101.012514} {\bibfield  {journal} {\bibinfo
   {journal} {Physical Review A}\ }\textbf {\bibinfo {volume} {101}},\ \bibinfo
  {pages} {012514} (\bibinfo {year} {2020})}\BibitemShut {NoStop}%
\bibitem [{\citenamefont {Indelicato}\ \emph {et~al.}(2007)\citenamefont
  {Indelicato}, \citenamefont {Santos}, \citenamefont {Boucard},\ and\
  \citenamefont {Desclaux}}]{2007_IndelicatoP_EurPhysJD}%
  \BibitemOpen
  \bibfield  {author} {\bibinfo {author} {\bibfnamefont {P.}~\bibnamefont
  {Indelicato}}, \bibinfo {author} {\bibfnamefont {J.~P.}\ \bibnamefont
  {Santos}}, \bibinfo {author} {\bibfnamefont {S.}~\bibnamefont {Boucard}},\
  and\ \bibinfo {author} {\bibfnamefont {J.-P.}\ \bibnamefont {Desclaux}},\
  }\href {https://doi.org/10.1140/epjd/e2007-00229-y} {\bibfield  {journal}
  {\bibinfo  {journal} {The European Physical Journal D}\ }\textbf {\bibinfo
  {volume} {45}},\ \bibinfo {pages} {155} (\bibinfo {year} {2007})}\BibitemShut
  {NoStop}%
\bibitem [{\citenamefont {Eliav}\ \emph {et~al.}(2015)\citenamefont {Eliav},
  \citenamefont {Fritzsche},\ and\ \citenamefont
  {Kaldor}}]{2015_EliavE_NuclPhysA}%
  \BibitemOpen
  \bibfield  {author} {\bibinfo {author} {\bibfnamefont {E.}~\bibnamefont
  {Eliav}}, \bibinfo {author} {\bibfnamefont {S.}~\bibnamefont {Fritzsche}},\
  and\ \bibinfo {author} {\bibfnamefont {U.}~\bibnamefont {Kaldor}},\ }\href
  {https://doi.org/10.1016/j.nuclphysa.2015.06.017} {\bibfield  {journal}
  {\bibinfo  {journal} {Nuclear Physics A}\ }\bibinfo {series} {Special
  {{Issue}} on {{Superheavy Elements}}},\ \textbf {\bibinfo {volume} {944}},\
  \bibinfo {pages} {518} (\bibinfo {year} {2015})}\BibitemShut {NoStop}%
\bibitem [{\citenamefont {Pershina}(2015)}]{2015_PershinaV_NuclPhysA}%
  \BibitemOpen
  \bibfield  {author} {\bibinfo {author} {\bibfnamefont {V.}~\bibnamefont
  {Pershina}},\ }\href {https://doi.org/10.1016/j.nuclphysa.2015.04.007}
  {\bibfield  {journal} {\bibinfo  {journal} {Nuclear Physics A}\ }\bibinfo
  {series} {Special {{Issue}} on {{Superheavy Elements}}},\ \textbf {\bibinfo
  {volume} {944}},\ \bibinfo {pages} {578} (\bibinfo {year}
  {2015})}\BibitemShut {NoStop}%
\bibitem [{\citenamefont {Schwerdtfeger}\ \emph {et~al.}(2015)\citenamefont
  {Schwerdtfeger}, \citenamefont {Pa{\v s}teka}, \citenamefont {Punnett},\ and\
  \citenamefont {Bowman}}]{2015_SchwerdtfegerP_NuclPhysA}%
  \BibitemOpen
  \bibfield  {author} {\bibinfo {author} {\bibfnamefont {P.}~\bibnamefont
  {Schwerdtfeger}}, \bibinfo {author} {\bibfnamefont {L.~F.}\ \bibnamefont
  {Pa{\v s}teka}}, \bibinfo {author} {\bibfnamefont {A.}~\bibnamefont
  {Punnett}},\ and\ \bibinfo {author} {\bibfnamefont {P.~O.}\ \bibnamefont
  {Bowman}},\ }\href {https://doi.org/10.1016/j.nuclphysa.2015.02.005}
  {\bibfield  {journal} {\bibinfo  {journal} {Nuclear Physics A}\ }\bibinfo
  {series} {Special {{Issue}} on {{Superheavy Elements}}},\ \textbf {\bibinfo
  {volume} {944}},\ \bibinfo {pages} {551} (\bibinfo {year}
  {2015})}\BibitemShut {NoStop}%
\bibitem [{\citenamefont {Pershina}(2019)}]{2019_PershinaV_RadChimActa}%
  \BibitemOpen
  \bibfield  {author} {\bibinfo {author} {\bibfnamefont {V.}~\bibnamefont
  {Pershina}},\ }\href {https://doi.org/10.1515/ract-2018-3098} {\bibfield
  {journal} {\bibinfo  {journal} {Radiochimica Acta}\ }\textbf {\bibinfo
  {volume} {107}},\ \bibinfo {pages} {833} (\bibinfo {year}
  {2019})}\BibitemShut {NoStop}%
\bibitem [{\citenamefont {Mosyagin}\ \emph {et~al.}(2020)\citenamefont
  {Mosyagin}, \citenamefont {Zaitsevskii},\ and\ \citenamefont
  {Titov}}]{2020_MosyaginN_IntJQuantChem}%
  \BibitemOpen
  \bibfield  {author} {\bibinfo {author} {\bibfnamefont {N.~S.}\ \bibnamefont
  {Mosyagin}}, \bibinfo {author} {\bibfnamefont {A.~V.}\ \bibnamefont
  {Zaitsevskii}},\ and\ \bibinfo {author} {\bibfnamefont {A.~V.}\ \bibnamefont
  {Titov}},\ }\href {https://doi.org/10.1002/qua.26076} {\bibfield  {journal}
  {\bibinfo  {journal} {International Journal of Quantum Chemistry}\ }\textbf
  {\bibinfo {volume} {120}},\ \bibinfo {pages} {e26076} (\bibinfo {year}
  {2020})}\BibitemShut {NoStop}%
\bibitem [{\citenamefont {Oganessian}\ and\ \citenamefont
  {Utyonkov}(2015)}]{2015_OganessianY_RepProgPhys}%
  \BibitemOpen
  \bibfield  {author} {\bibinfo {author} {\bibfnamefont {Y.~T.}\ \bibnamefont
  {Oganessian}}\ and\ \bibinfo {author} {\bibfnamefont {V.~K.}\ \bibnamefont
  {Utyonkov}},\ }\href {https://doi.org/10.1088/0034-4885/78/3/036301}
  {\bibfield  {journal} {\bibinfo  {journal} {Reports on Progress in Physics}\
  }\textbf {\bibinfo {volume} {78}},\ \bibinfo {pages} {036301} (\bibinfo
  {year} {2015})}\BibitemShut {NoStop}%
\bibitem [{\citenamefont {Oganessian}\ \emph {et~al.}(2017)\citenamefont
  {Oganessian}, \citenamefont {Sobiczewski},\ and\ \citenamefont
  {{Ter-Akopian}}}]{2017_OganessianY_PhysScrip}%
  \BibitemOpen
  \bibfield  {author} {\bibinfo {author} {\bibfnamefont {Y.~T.}\ \bibnamefont
  {Oganessian}}, \bibinfo {author} {\bibfnamefont {A.}~\bibnamefont
  {Sobiczewski}},\ and\ \bibinfo {author} {\bibfnamefont {G.~M.}\ \bibnamefont
  {{Ter-Akopian}}},\ }\href {https://doi.org/10.1088/1402-4896/aa53c1}
  {\bibfield  {journal} {\bibinfo  {journal} {Physica Scripta}\ }\textbf
  {\bibinfo {volume} {92}},\ \bibinfo {pages} {023003} (\bibinfo {year}
  {2017})}\BibitemShut {NoStop}%
\bibitem [{\citenamefont {Giuliani}\ \emph {et~al.}(2019)\citenamefont
  {Giuliani}, \citenamefont {Matheson}, \citenamefont {Nazarewicz},
  \citenamefont {Olsen}, \citenamefont {Reinhard}, \citenamefont {Sadhukhan},
  \citenamefont {Schuetrumpf}, \citenamefont {Schunck},\ and\ \citenamefont
  {Schwerdtfeger}}]{2019_GiulianiS_RevModPhys}%
  \BibitemOpen
  \bibfield  {author} {\bibinfo {author} {\bibfnamefont {S.~A.}\ \bibnamefont
  {Giuliani}}, \bibinfo {author} {\bibfnamefont {Z.}~\bibnamefont {Matheson}},
  \bibinfo {author} {\bibfnamefont {W.}~\bibnamefont {Nazarewicz}}, \bibinfo
  {author} {\bibfnamefont {E.}~\bibnamefont {Olsen}}, \bibinfo {author}
  {\bibfnamefont {P.-G.}\ \bibnamefont {Reinhard}}, \bibinfo {author}
  {\bibfnamefont {J.}~\bibnamefont {Sadhukhan}}, \bibinfo {author}
  {\bibfnamefont {B.}~\bibnamefont {Schuetrumpf}}, \bibinfo {author}
  {\bibfnamefont {N.}~\bibnamefont {Schunck}},\ and\ \bibinfo {author}
  {\bibfnamefont {P.}~\bibnamefont {Schwerdtfeger}},\ }\href
  {https://doi.org/10.1103/RevModPhys.91.011001} {\bibfield  {journal}
  {\bibinfo  {journal} {Reviews of Modern Physics}\ }\textbf {\bibinfo {volume}
  {91}},\ \bibinfo {pages} {011001} (\bibinfo {year} {2019})}\BibitemShut
  {NoStop}%
\bibitem [{\citenamefont {Eliav}\ \emph {et~al.}(1996)\citenamefont {Eliav},
  \citenamefont {Kaldor}, \citenamefont {Ishikawa},\ and\ \citenamefont
  {Pyykk{\"o}}}]{1996_EliavE_PhysRevLett}%
  \BibitemOpen
  \bibfield  {author} {\bibinfo {author} {\bibfnamefont {E.}~\bibnamefont
  {Eliav}}, \bibinfo {author} {\bibfnamefont {U.}~\bibnamefont {Kaldor}},
  \bibinfo {author} {\bibfnamefont {Y.}~\bibnamefont {Ishikawa}},\ and\
  \bibinfo {author} {\bibfnamefont {P.}~\bibnamefont {Pyykk{\"o}}},\ }\href
  {https://doi.org/10.1103/PhysRevLett.77.5350} {\bibfield  {journal} {\bibinfo
   {journal} {Physical Review Letters}\ }\textbf {\bibinfo {volume} {77}},\
  \bibinfo {pages} {5350} (\bibinfo {year} {1996})}\BibitemShut {NoStop}%
\bibitem [{\citenamefont {Goidenko}\ \emph {et~al.}(2003)\citenamefont
  {Goidenko}, \citenamefont {Labzowsky}, \citenamefont {Eliav}, \citenamefont
  {Kaldor},\ and\ \citenamefont {Pyykk{\"o}}}]{2003_GoidenkoI_PhysRevA}%
  \BibitemOpen
  \bibfield  {author} {\bibinfo {author} {\bibfnamefont {I.}~\bibnamefont
  {Goidenko}}, \bibinfo {author} {\bibfnamefont {L.}~\bibnamefont {Labzowsky}},
  \bibinfo {author} {\bibfnamefont {E.}~\bibnamefont {Eliav}}, \bibinfo
  {author} {\bibfnamefont {U.}~\bibnamefont {Kaldor}},\ and\ \bibinfo {author}
  {\bibfnamefont {P.}~\bibnamefont {Pyykk{\"o}}},\ }\href
  {https://doi.org/10.1103/PhysRevA.67.020102} {\bibfield  {journal} {\bibinfo
  {journal} {Physical Review A}\ }\textbf {\bibinfo {volume} {67}},\ \bibinfo
  {pages} {020102} (\bibinfo {year} {2003})}\BibitemShut {NoStop}%
\bibitem [{\citenamefont {Lackenby}\ \emph {et~al.}(2018)\citenamefont
  {Lackenby}, \citenamefont {Dzuba},\ and\ \citenamefont
  {Flambaum}}]{2018_LackenbyB_PhysRevA98_Og}%
  \BibitemOpen
  \bibfield  {author} {\bibinfo {author} {\bibfnamefont {B.~G.~C.}\
  \bibnamefont {Lackenby}}, \bibinfo {author} {\bibfnamefont {V.~A.}\
  \bibnamefont {Dzuba}},\ and\ \bibinfo {author} {\bibfnamefont {V.~V.}\
  \bibnamefont {Flambaum}},\ }\href
  {https://doi.org/10.1103/PhysRevA.98.042512} {\bibfield  {journal} {\bibinfo
  {journal} {Physical Review A}\ }\textbf {\bibinfo {volume} {98}},\ \bibinfo
  {pages} {042512} (\bibinfo {year} {2018})}\BibitemShut {NoStop}%
\bibitem [{\citenamefont {Saue}\ \emph {et~al.}(2020)\citenamefont {Saue},
  \citenamefont {Bast}, \citenamefont {Gomes}, \citenamefont {Jensen},
  \citenamefont {Visscher}, \citenamefont {Aucar}, \citenamefont {Di~Remigio},
  \citenamefont {Dyall}, \citenamefont {Eliav}, \citenamefont {Fasshauer},
  \citenamefont {Fleig}, \citenamefont {Halbert}, \citenamefont {Hedeg{\aa}rd},
  \citenamefont {{Helmich-Paris}}, \citenamefont {Ilia{\v s}}, \citenamefont
  {Jacob}, \citenamefont {Knecht}, \citenamefont {Laerdahl}, \citenamefont
  {Vidal}, \citenamefont {Nayak}, \citenamefont {Olejniczak}, \citenamefont
  {Olsen}, \citenamefont {Pernpointner}, \citenamefont {Senjean}, \citenamefont
  {Shee}, \citenamefont {Sunaga},\ and\ \citenamefont {{van
  Stralen}}}]{2020_SaueT_JChemPhys}%
  \BibitemOpen
  \bibfield  {author} {\bibinfo {author} {\bibfnamefont {T.}~\bibnamefont
  {Saue}}, \bibinfo {author} {\bibfnamefont {R.}~\bibnamefont {Bast}}, \bibinfo
  {author} {\bibfnamefont {A.~S.~P.}\ \bibnamefont {Gomes}}, \bibinfo {author}
  {\bibfnamefont {H.~J.~A.}\ \bibnamefont {Jensen}}, \bibinfo {author}
  {\bibfnamefont {L.}~\bibnamefont {Visscher}}, \bibinfo {author}
  {\bibfnamefont {I.~A.}\ \bibnamefont {Aucar}}, \bibinfo {author}
  {\bibfnamefont {R.}~\bibnamefont {Di~Remigio}}, \bibinfo {author}
  {\bibfnamefont {K.~G.}\ \bibnamefont {Dyall}}, \bibinfo {author}
  {\bibfnamefont {E.}~\bibnamefont {Eliav}}, \bibinfo {author} {\bibfnamefont
  {E.}~\bibnamefont {Fasshauer}}, \bibinfo {author} {\bibfnamefont
  {T.}~\bibnamefont {Fleig}}, \bibinfo {author} {\bibfnamefont
  {L.}~\bibnamefont {Halbert}}, \bibinfo {author} {\bibfnamefont {E.~D.}\
  \bibnamefont {Hedeg{\aa}rd}}, \bibinfo {author} {\bibfnamefont
  {B.}~\bibnamefont {{Helmich-Paris}}}, \bibinfo {author} {\bibfnamefont
  {M.}~\bibnamefont {Ilia{\v s}}}, \bibinfo {author} {\bibfnamefont {C.~R.}\
  \bibnamefont {Jacob}}, \bibinfo {author} {\bibfnamefont {S.}~\bibnamefont
  {Knecht}}, \bibinfo {author} {\bibfnamefont {J.~K.}\ \bibnamefont
  {Laerdahl}}, \bibinfo {author} {\bibfnamefont {M.~L.}\ \bibnamefont {Vidal}},
  \bibinfo {author} {\bibfnamefont {M.~K.}\ \bibnamefont {Nayak}}, \bibinfo
  {author} {\bibfnamefont {M.}~\bibnamefont {Olejniczak}}, \bibinfo {author}
  {\bibfnamefont {J.~M.~H.}\ \bibnamefont {Olsen}}, \bibinfo {author}
  {\bibfnamefont {M.}~\bibnamefont {Pernpointner}}, \bibinfo {author}
  {\bibfnamefont {B.}~\bibnamefont {Senjean}}, \bibinfo {author} {\bibfnamefont
  {A.}~\bibnamefont {Shee}}, \bibinfo {author} {\bibfnamefont {A.}~\bibnamefont
  {Sunaga}},\ and\ \bibinfo {author} {\bibfnamefont {J.~N.~P.}\ \bibnamefont
  {{van Stralen}}},\ }\href {https://doi.org/10.1063/5.0004844} {\bibfield
  {journal} {\bibinfo  {journal} {The Journal of Chemical Physics}\ }\textbf
  {\bibinfo {volume} {152}},\ \bibinfo {pages} {204104} (\bibinfo {year}
  {2020})}\BibitemShut {NoStop}%
\bibitem [{\citenamefont {Oleynichenko}\ \emph {et~al.}(2020)\citenamefont
  {Oleynichenko}, \citenamefont {Zaitsevskii}, \citenamefont {Skripnikov},\
  and\ \citenamefont {Eliav}}]{2020_OleynichenkoA_Symmetry}%
  \BibitemOpen
  \bibfield  {author} {\bibinfo {author} {\bibfnamefont {A.~V.}\ \bibnamefont
  {Oleynichenko}}, \bibinfo {author} {\bibfnamefont {A.}~\bibnamefont
  {Zaitsevskii}}, \bibinfo {author} {\bibfnamefont {L.~V.}\ \bibnamefont
  {Skripnikov}},\ and\ \bibinfo {author} {\bibfnamefont {E.}~\bibnamefont
  {Eliav}},\ }\href {https://doi.org/10.3390/sym12071101} {\bibfield  {journal}
  {\bibinfo  {journal} {Symmetry}\ }\textbf {\bibinfo {volume} {12}},\ \bibinfo
  {pages} {1101} (\bibinfo {year} {2020})}\BibitemShut {NoStop}%
\bibitem [{\citenamefont {Tupitsyn}\ and\ \citenamefont
  {Loginov}(2003)}]{2003_TupitsynI_OptSpectrosc}%
  \BibitemOpen
  \bibfield  {author} {\bibinfo {author} {\bibfnamefont {I.~I.}\ \bibnamefont
  {Tupitsyn}}\ and\ \bibinfo {author} {\bibfnamefont {A.~V.}\ \bibnamefont
  {Loginov}},\ }\href {https://doi.org/10.1134/1.1563671} {\bibfield  {journal}
  {\bibinfo  {journal} {Optics and Spectroscopy}\ }\textbf {\bibinfo {volume}
  {94}},\ \bibinfo {pages} {319} (\bibinfo {year} {2003})}\BibitemShut
  {NoStop}%
\bibitem [{\citenamefont {Tupitsyn}\ \emph {et~al.}(2003)\citenamefont
  {Tupitsyn}, \citenamefont {Shabaev}, \citenamefont {{Crespo
  L{\'o}pez-Urrutia}}, \citenamefont {Dragani{\'c}}, \citenamefont
  {Soria~Orts},\ and\ \citenamefont {Ullrich}}]{2003_TupitsynI_PhysRevA}%
  \BibitemOpen
  \bibfield  {author} {\bibinfo {author} {\bibfnamefont {I.~I.}\ \bibnamefont
  {Tupitsyn}}, \bibinfo {author} {\bibfnamefont {V.~M.}\ \bibnamefont
  {Shabaev}}, \bibinfo {author} {\bibfnamefont {J.~R.}\ \bibnamefont {{Crespo
  L{\'o}pez-Urrutia}}}, \bibinfo {author} {\bibfnamefont {I.}~\bibnamefont
  {Dragani{\'c}}}, \bibinfo {author} {\bibfnamefont {R.}~\bibnamefont
  {Soria~Orts}},\ and\ \bibinfo {author} {\bibfnamefont {J.}~\bibnamefont
  {Ullrich}},\ }\href {https://doi.org/10.1103/PhysRevA.68.022511} {\bibfield
  {journal} {\bibinfo  {journal} {Physical Review A}\ }\textbf {\bibinfo
  {volume} {68}},\ \bibinfo {pages} {022511} (\bibinfo {year}
  {2003})}\BibitemShut {NoStop}%
\bibitem [{\citenamefont {Tupitsyn}\ \emph {et~al.}(2005)\citenamefont
  {Tupitsyn}, \citenamefont {Volotka}, \citenamefont {Glazov}, \citenamefont
  {Shabaev}, \citenamefont {Plunien}, \citenamefont {{Crespo
  L{\'o}pez-Urrutia}}, \citenamefont {Lapierre},\ and\ \citenamefont
  {Ullrich}}]{2005_TupitsynI_PhysRevA}%
  \BibitemOpen
  \bibfield  {author} {\bibinfo {author} {\bibfnamefont {I.~I.}\ \bibnamefont
  {Tupitsyn}}, \bibinfo {author} {\bibfnamefont {A.~V.}\ \bibnamefont
  {Volotka}}, \bibinfo {author} {\bibfnamefont {D.~A.}\ \bibnamefont {Glazov}},
  \bibinfo {author} {\bibfnamefont {V.~M.}\ \bibnamefont {Shabaev}}, \bibinfo
  {author} {\bibfnamefont {G.}~\bibnamefont {Plunien}}, \bibinfo {author}
  {\bibfnamefont {J.~R.}\ \bibnamefont {{Crespo L{\'o}pez-Urrutia}}}, \bibinfo
  {author} {\bibfnamefont {A.}~\bibnamefont {Lapierre}},\ and\ \bibinfo
  {author} {\bibfnamefont {J.}~\bibnamefont {Ullrich}},\ }\href
  {https://doi.org/10.1103/PhysRevA.72.062503} {\bibfield  {journal} {\bibinfo
  {journal} {Physical Review A}\ }\textbf {\bibinfo {volume} {72}},\ \bibinfo
  {pages} {062503} (\bibinfo {year} {2005})}\BibitemShut {NoStop}%
\bibitem [{\citenamefont {Tupitsyn}\ \emph {et~al.}(2018)\citenamefont
  {Tupitsyn}, \citenamefont {Zubova}, \citenamefont {Shabaev}, \citenamefont
  {Plunien},\ and\ \citenamefont {St{\"o}hlker}}]{2018_TupitsynI_PhysRevA}%
  \BibitemOpen
  \bibfield  {author} {\bibinfo {author} {\bibfnamefont {I.~I.}\ \bibnamefont
  {Tupitsyn}}, \bibinfo {author} {\bibfnamefont {N.~A.}\ \bibnamefont
  {Zubova}}, \bibinfo {author} {\bibfnamefont {V.~M.}\ \bibnamefont {Shabaev}},
  \bibinfo {author} {\bibfnamefont {G.}~\bibnamefont {Plunien}},\ and\ \bibinfo
  {author} {\bibfnamefont {T.}~\bibnamefont {St{\"o}hlker}},\ }\href
  {https://doi.org/10.1103/PhysRevA.98.022517} {\bibfield  {journal} {\bibinfo
  {journal} {Physical Review A}\ }\textbf {\bibinfo {volume} {98}},\ \bibinfo
  {pages} {022517} (\bibinfo {year} {2018})}\BibitemShut {NoStop}%
\bibitem [{\citenamefont {Sikkema}\ \emph {et~al.}(2009)\citenamefont
  {Sikkema}, \citenamefont {Visscher}, \citenamefont {Saue},\ and\
  \citenamefont {Ilia{\v s}}}]{2009_SikkemaJ_JChemPhys}%
  \BibitemOpen
  \bibfield  {author} {\bibinfo {author} {\bibfnamefont {J.}~\bibnamefont
  {Sikkema}}, \bibinfo {author} {\bibfnamefont {L.}~\bibnamefont {Visscher}},
  \bibinfo {author} {\bibfnamefont {T.}~\bibnamefont {Saue}},\ and\ \bibinfo
  {author} {\bibfnamefont {M.}~\bibnamefont {Ilia{\v s}}},\ }\href
  {https://doi.org/10.1063/1.3239505} {\bibfield  {journal} {\bibinfo
  {journal} {The Journal of Chemical Physics}\ }\textbf {\bibinfo {volume}
  {131}},\ \bibinfo {pages} {124116} (\bibinfo {year} {2009})}\BibitemShut
  {NoStop}%
\bibitem [{\citenamefont {Norcross}(1973)}]{1973_NorcrossD_PhysRevA}%
  \BibitemOpen
  \bibfield  {author} {\bibinfo {author} {\bibfnamefont {D.~W.}\ \bibnamefont
  {Norcross}},\ }\href {https://doi.org/10.1103/PhysRevA.7.606} {\bibfield
  {journal} {\bibinfo  {journal} {Physical Review A}\ }\textbf {\bibinfo
  {volume} {7}},\ \bibinfo {pages} {606} (\bibinfo {year} {1973})}\BibitemShut
  {NoStop}%
\bibitem [{\citenamefont {Dalgarno}\ \emph {et~al.}(1970)\citenamefont
  {Dalgarno}, \citenamefont {Bottcher},\ and\ \citenamefont
  {Victor}}]{1970_DalgarnoA_ChemPhysLett}%
  \BibitemOpen
  \bibfield  {author} {\bibinfo {author} {\bibfnamefont {A.}~\bibnamefont
  {Dalgarno}}, \bibinfo {author} {\bibfnamefont {C.}~\bibnamefont {Bottcher}},\
  and\ \bibinfo {author} {\bibfnamefont {G.~A.}\ \bibnamefont {Victor}},\
  }\href {https://doi.org/10.1016/0009-2614(70)80304-9} {\bibfield  {journal}
  {\bibinfo  {journal} {Chemical Physics Letters}\ }\textbf {\bibinfo {volume}
  {7}},\ \bibinfo {pages} {265} (\bibinfo {year} {1970})}\BibitemShut {NoStop}%
\bibitem [{\citenamefont {Baylis}(1977)}]{1977_BaylisW_JPhysBAtomMolPhys}%
  \BibitemOpen
  \bibfield  {author} {\bibinfo {author} {\bibfnamefont {W.~E.}\ \bibnamefont
  {Baylis}},\ }\href {https://doi.org/10.1088/0022-3700/10/16/001} {\bibfield
  {journal} {\bibinfo  {journal} {Journal of Physics B: Atomic, Molecular and
  Optical Physics}\ }\textbf {\bibinfo {volume} {10}},\ \bibinfo {pages} {L583}
  (\bibinfo {year} {1977})}\BibitemShut {NoStop}%
\bibitem [{\citenamefont {Mitroy}\ \emph {et~al.}(2010)\citenamefont {Mitroy},
  \citenamefont {Safronova},\ and\ \citenamefont
  {Clark}}]{2010_MitroyJ_JPhysB}%
  \BibitemOpen
  \bibfield  {author} {\bibinfo {author} {\bibfnamefont {J.}~\bibnamefont
  {Mitroy}}, \bibinfo {author} {\bibfnamefont {M.~S.}\ \bibnamefont
  {Safronova}},\ and\ \bibinfo {author} {\bibfnamefont {C.~W.}\ \bibnamefont
  {Clark}},\ }\href {https://doi.org/10.1088/0953-4075/43/20/202001} {\bibfield
   {journal} {\bibinfo  {journal} {Journal of Physics B: Atomic, Molecular and
  Optical Physics}\ }\textbf {\bibinfo {volume} {43}},\ \bibinfo {pages}
  {202001} (\bibinfo {year} {2010})}\BibitemShut {NoStop}%
\bibitem [{\citenamefont {Bates}\ and\ \citenamefont
  {Massey}(1943)}]{1943_BatesD}%
  \BibitemOpen
  \bibfield  {author} {\bibinfo {author} {\bibfnamefont {D.~R.}\ \bibnamefont
  {Bates}}\ and\ \bibinfo {author} {\bibfnamefont {H.~S.~W.}\ \bibnamefont
  {Massey}},\ }\href {https://doi.org/10.1098/rsta.1943.0001} {\bibfield
  {journal} {\bibinfo  {journal} {Philosophical Transactions of the Royal
  Society of London. Series A, Mathematical and Physical Sciences}\ }\textbf
  {\bibinfo {volume} {239}},\ \bibinfo {pages} {269} (\bibinfo {year}
  {1943})}\BibitemShut {NoStop}%
\bibitem [{\citenamefont {Pyykk{\"o}}\ and\ \citenamefont
  {Zhao}(2003)}]{2003_PyykkoP_JPhysB}%
  \BibitemOpen
  \bibfield  {author} {\bibinfo {author} {\bibfnamefont {P.}~\bibnamefont
  {Pyykk{\"o}}}\ and\ \bibinfo {author} {\bibfnamefont {L.-B.}\ \bibnamefont
  {Zhao}},\ }\href {https://doi.org/10.1088/0953-4075/36/8/302} {\bibfield
  {journal} {\bibinfo  {journal} {Journal of Physics B: Atomic, Molecular and
  Optical Physics}\ }\textbf {\bibinfo {volume} {36}},\ \bibinfo {pages} {1469}
  (\bibinfo {year} {2003})}\BibitemShut {NoStop}%
\bibitem [{\citenamefont {Flambaum}\ and\ \citenamefont
  {Ginges}(2005)}]{2005_FlambaumV_PhysRevA}%
  \BibitemOpen
  \bibfield  {author} {\bibinfo {author} {\bibfnamefont {V.~V.}\ \bibnamefont
  {Flambaum}}\ and\ \bibinfo {author} {\bibfnamefont {J.~S.~M.}\ \bibnamefont
  {Ginges}},\ }\href {https://doi.org/10.1103/PhysRevA.72.052115} {\bibfield
  {journal} {\bibinfo  {journal} {Physical Review A}\ }\textbf {\bibinfo
  {volume} {72}},\ \bibinfo {pages} {052115} (\bibinfo {year}
  {2005})}\BibitemShut {NoStop}%
\bibitem [{\citenamefont {Tupitsyn}\ and\ \citenamefont
  {Berseneva}(2013)}]{2013_TupitsynI_OptSpectrosc}%
  \BibitemOpen
  \bibfield  {author} {\bibinfo {author} {\bibfnamefont {I.~I.}\ \bibnamefont
  {Tupitsyn}}\ and\ \bibinfo {author} {\bibfnamefont {E.~V.}\ \bibnamefont
  {Berseneva}},\ }\href {https://doi.org/10.1134/S0030400X13050214} {\bibfield
  {journal} {\bibinfo  {journal} {Optics and Spectroscopy}\ }\textbf {\bibinfo
  {volume} {114}},\ \bibinfo {pages} {682} (\bibinfo {year}
  {2013})}\BibitemShut {NoStop}%
\bibitem [{\citenamefont {Shabaev}\ \emph {et~al.}(2013)\citenamefont
  {Shabaev}, \citenamefont {Tupitsyn},\ and\ \citenamefont
  {Yerokhin}}]{2013_ShabaevV_PhysRevA}%
  \BibitemOpen
  \bibfield  {author} {\bibinfo {author} {\bibfnamefont {V.~M.}\ \bibnamefont
  {Shabaev}}, \bibinfo {author} {\bibfnamefont {I.~I.}\ \bibnamefont
  {Tupitsyn}},\ and\ \bibinfo {author} {\bibfnamefont {V.~A.}\ \bibnamefont
  {Yerokhin}},\ }\href {https://doi.org/10.1103/PhysRevA.88.012513} {\bibfield
  {journal} {\bibinfo  {journal} {Physical Review A}\ }\textbf {\bibinfo
  {volume} {88}},\ \bibinfo {pages} {012513} (\bibinfo {year}
  {2013})}\BibitemShut {NoStop}%
\bibitem [{\citenamefont {Shabaev}\ \emph {et~al.}(2015)\citenamefont
  {Shabaev}, \citenamefont {Tupitsyn},\ and\ \citenamefont
  {Yerokhin}}]{2015_ShabaevV_CompPhysComm}%
  \BibitemOpen
  \bibfield  {author} {\bibinfo {author} {\bibfnamefont {V.~M.}\ \bibnamefont
  {Shabaev}}, \bibinfo {author} {\bibfnamefont {I.~I.}\ \bibnamefont
  {Tupitsyn}},\ and\ \bibinfo {author} {\bibfnamefont {V.~A.}\ \bibnamefont
  {Yerokhin}},\ }\href {https://doi.org/10.1016/j.cpc.2014.12.002} {\bibfield
  {journal} {\bibinfo  {journal} {Computer Physics Communications}\ }\textbf
  {\bibinfo {volume} {189}},\ \bibinfo {pages} {175} (\bibinfo {year}
  {2015})}\BibitemShut {NoStop}%
\bibitem [{\citenamefont {Shabaev}\ \emph {et~al.}(2018)\citenamefont
  {Shabaev}, \citenamefont {Tupitsyn},\ and\ \citenamefont
  {Yerokhin}}]{2018_ShabaevV_CompPhysComm}%
  \BibitemOpen
  \bibfield  {author} {\bibinfo {author} {\bibfnamefont {V.~M.}\ \bibnamefont
  {Shabaev}}, \bibinfo {author} {\bibfnamefont {I.~I.}\ \bibnamefont
  {Tupitsyn}},\ and\ \bibinfo {author} {\bibfnamefont {V.~A.}\ \bibnamefont
  {Yerokhin}},\ }\href {https://doi.org/10.1016/j.cpc.2017.10.007} {\bibfield
  {journal} {\bibinfo  {journal} {Computer Physics Communications}\ }\textbf
  {\bibinfo {volume} {223}},\ \bibinfo {pages} {69} (\bibinfo {year}
  {2018})}\BibitemShut {NoStop}%
\bibitem [{\citenamefont {Mosyagin}\ \emph {et~al.}(2000)\citenamefont
  {Mosyagin}, \citenamefont {Eliav}, \citenamefont {Titov},\ and\ \citenamefont
  {Kaldor}}]{Mosyagin:00}%
  \BibitemOpen
  \bibfield  {author} {\bibinfo {author} {\bibfnamefont {N.~S.}\ \bibnamefont
  {Mosyagin}}, \bibinfo {author} {\bibfnamefont {E.}~\bibnamefont {Eliav}},
  \bibinfo {author} {\bibfnamefont {A.~V.}\ \bibnamefont {Titov}},\ and\
  \bibinfo {author} {\bibfnamefont {U.}~\bibnamefont {Kaldor}},\ }\href@noop {}
  {\bibfield  {journal} {\bibinfo  {journal} {J. Phys. B}\ }\textbf {\bibinfo
  {volume} {33}},\ \bibinfo {pages} {667} (\bibinfo {year} {2000})}\BibitemShut
  {NoStop}%
\bibitem [{\citenamefont {Skripnikov}(2021)}]{Skripnikov:2021}%
  \BibitemOpen
  \bibfield  {author} {\bibinfo {author} {\bibfnamefont {L.~V.}\ \bibnamefont
  {Skripnikov}},\ }\href@noop {} {\bibfield  {journal} {\bibinfo  {journal} {J.
  Chem. Phys., in press}\ } (\bibinfo {year} {2021})}\BibitemShut {NoStop}%
\end{thebibliography}%

\appendix
\newpage
\section{FSCC basis optimization}\label{seq:appendix_a}
We begin the optimization of the Dyall's basis set AAE4Z by adding new functions one-by-one to the set starting from the $s$-type basis functions. 
We vary the parameters of the added functions tracking the change of EA in the FSCC-SD calculations.
To save the time and computational resources, we apply three simplifications of the problem during the basis-set optimization procedure: (i) the GRECP operator is used, which for the case of Og replaces 92 electrons with their pseudopotential leaving us overall with a problem for the $6s^26p^66d^{10}7s^27p^6$ configuration; (ii) the FSCC-SD correlated electrons are $6s6p6d7s7p$ and AS includes virtual DF orbitals with energies $\varepsilon^{\mathrm{DF}} < 40$ a.u.; (iii) the basis functions with the orbital quantum number $L>3$ are removed from the basis set and are to be optimized later.
\par
The optimization procedure is similar to that proposed in Ref.~\cite{Mosyagin:00} and is as follows.
Let us denote the initial basis set we are working with as $\chi_0$.
We add the $s$-type basis function with a parameter $\zeta$, $\gamma^s_1(\zeta)$, to the set $\chi_0$ and perform the calculations using the FSCC-SD method for some range of the parameter $\zeta \in [\zeta_{\mathrm{min}}; \zeta_{\mathrm{max}}]$.
The basis function $\gamma^s_1(\zeta_1)$ which inclusion to the set leads to the largest shift of EA, compared to the related result without this function, is permanently incorporated into the basis set, $\chi_0 + \gamma^s_1(\zeta_1) \equiv \chi^s_1$.
The procedure continues unless the change of EA with the addition of $k+1$-th basis function to the basis set $\chi^s_k$ is less than the acceptable uncertainty of EA. 
The resulted basis set now contains the $k$ additional optimized basis functions of the $s$-type and is denoted as $\chi_0 + \sum^k_{m=1} \gamma(\zeta_m) \equiv \chi^s$.
The same procedure is repeated for the $p$-type basis functions, while the parameters of the basis functions from the previous stage remain unchanged.
The basis set with the $k$ optimized $s$- and $k'$ optimized $p$-type basis functions is now $\chi^s + \sum_{m=1}^{k'} \gamma^p_m(\zeta_m) \equiv \chi^{p}$.
The optimization procedure of the next $L$-type basis functions continues until the desired balance between the computational cost and the required accuracy is reached, leaving us with the optimized basis set denoted as $\chi^{L_{\mathrm{max}}}$.
\par
\begin{figure}[H]
  \centering
  \includegraphics[width=\columnwidth]{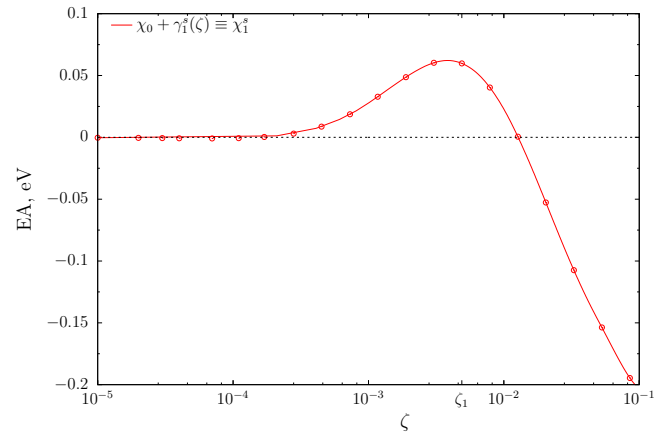}
  \caption{EA for Og calculated with the $H_{\mathrm{GRECP}}$ Hamiltonian by means of the FSCC-SD method exploiting the basis set~$\chi_0 + \gamma^s_1$ versus the parameter $\zeta$ of the basis function $\gamma^s_1(\zeta)$, in eV.}
  \label{fig:1}
\end{figure}
\begin{figure}[H]
  \centering
  \includegraphics[width=\linewidth]{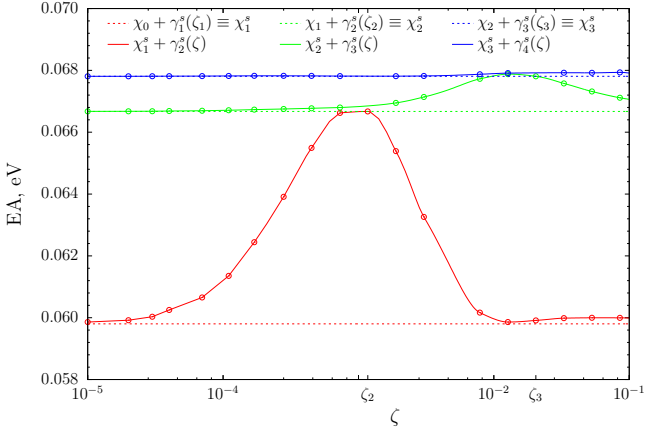}
  \caption{EA for Og calculated with the $H_{\mathrm{GRECP}}$ Hamiltonian by means of the FSCC-SD method for the subsequently enlarged basis sets $\chi^s_2$, $\chi^s_3$, and $\chi^s_4$ versus the parameter $\zeta$ of the basis functions $\gamma^s(\zeta)$, in eV.
  The basis set $\chi^s_1$ includes one optimized $s$-type basis function obtained on the previous stage of the optimization.}
  \label{fig:2}
\end{figure}

In Fig.~\ref{fig:1}, the very first step of the optimization procedure for the $s$-type basis functions is pictured.
The negative EA means that the addition of the localized function $\gamma^s_1(\zeta)$ with $\zeta \in [10^{-2};10^{-1}]$ contributes more to the correlation energy of the atom rather than the anion.
However, the addition of $\gamma^s_1(\zeta)$ with $\zeta \in [10^{-3};10^{-2}]$ to the initial basis set $\chi_0$ yields a bound state for Og$^-$ with a maximum positive EA $\approx0.060$~eV.
This corresponds to the increased quality of the basis set in the spatial region where the~$8s$~electron is localized.
With the addition of the basis functions with the parameter~$\zeta < 10^{-3}$ EA tends~to~$0$.
This is due to the fact that such delocalized functions are equally worthless for the correlations in the valence region for both atom and anion.
To sum up, according to the algorithm described above, $\gamma^s_1(\zeta)$ with $\zeta\equiv\zeta_1\approx4.89\cdot10^{-3}$ is added to $\chi_0$.
The complete basis-set optimization procedure for the $s$-type functions is shown in Fig.~\ref{fig:2}. 
The value of EA calculated with the basis set $\chi_1^s$ is shown with the red dashed line.
The value of EA evaluated with the additional basis function $\gamma^s_2(\zeta)$ is pictured with the red solid line.
The maximum deviation in EA for the basis set $\chi^s_1 + \gamma^s_2(\zeta)$ compared to the results obtained without the function $\gamma^s_2(\zeta)$ lies in the vicinity of $\zeta\equiv\zeta_2\approx1.17\cdot10^{-3}$ and indicates the further improvement of the $8s$ electron correlation energy in this spatial region.
The basis function with this parameter is permanently added to the basis set.
Proceeding, the green solid line shows the next step of the optimization and corresponds to the EA values calculated with the $\chi_2^s + \gamma^s_3(\zeta)$ basis set.
Compared to the $\chi^s_2$ value, the largest contribution to EA comes from the additional basis function $\gamma^s_3(\zeta)$ with $\zeta\equiv\zeta_3 \approx 2.04\cdot10^{-2}$.
The contribution to EA from $\gamma^s_3(\zeta_3)$ is several times smaller than that from $\gamma^s_2(\zeta_2)$.
At last, the blue solid line represents the contribution to EA, with $\gamma^s_4(\zeta)$ being added to the set $\chi_3^s$.
According to the calculations, this additional basis function has a negligible contribution to EA.
This contribution is included into the total uncertainty for EA.
Finally, we conclude that $\chi_3^s \equiv \chi^s$ contains the most important $s$-type basis functions needed for the calculation of EA in the FSCC-SD model.
One may argue that the parameters and the number of the basis functions $\gamma^s(\zeta)$ are not optimal since the initial basis $\chi_0$ does not include the diffuse $p$-, $d$-, $f$-type functions as well as any $g$- and higher-$L$-type basis functions.
We will turn back to this question below, when the optimal basis set which includes the optimized higher-$L$-type basis functions is constructed, and verify that adding the basis function $\gamma^s_4(\zeta)$ does not contribute to the EA value.
\par
\begin{figure}[H]
  \centering
  \includegraphics[width=\linewidth]{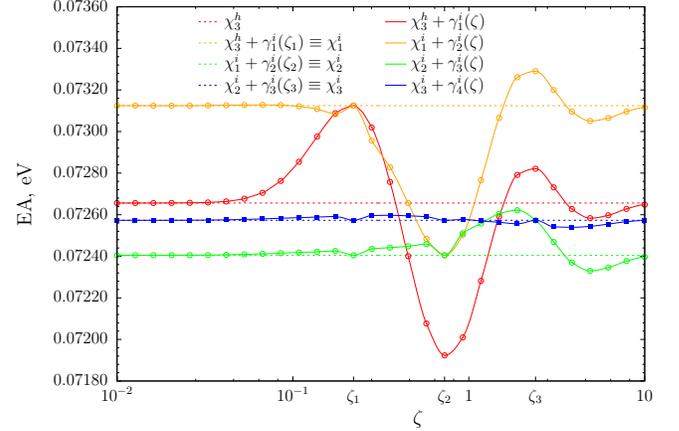}
  \caption{EA for Og calculated with the $H_{\mathrm{GRECP}}$ Hamiltonian by means of the FSCC-SD method for the subsequently enlarged basis sets $\chi^i_1$, $\chi^i_2$, $\chi^i_3$, and $\chi^i_4$ versus the parameter $\zeta$ of the basis functions $\gamma^i_k(\zeta),\,k=1,\dots,4$, in eV.
  The basis set $\chi^h$ includes the optimized functions for all $L$ up to $h$.}
  \label{fig:3}
\end{figure}

In Fig.~\ref{fig:3}, the final stage of the basis-set optimization procedure is presented.
The basis set $\chi^h$ which includes all optimized basis functions with $L$ up to $h$ is considered as the starting point.
The value of EA obtained in the $\chi^h$ basis set is shown with the red dashed line.
Adding $\gamma^i_1(\zeta)$ to the basis set $\chi^h$ results in several extremes for EA.
Here, it should be noted that there is an approximate additive property of the contributions to EA from various $\gamma^i(\zeta)$.
The calculations with the 3 optimized $i$-type basis functions, \textit{i.e.} with the basis set $\chi_3^i$, yields a contribution of $-0.00011$ eV to the value of EA, whereas the sum of the contributions to EA calculated with the basis functions $\gamma^i_1(\zeta_1)$, $\gamma^i_2(\zeta_2)$, and $\gamma^i_3(\zeta_3)$, taken separately, amounts to $-0.00009$ eV.
Since the change in the EA value with adding the $i$-type functions became comparable with the uncertainty, it was decided to stop further optimization of the basis set.
To ensure that the addition of the $g$-, $h$-, and $i$-type basis functions does not spoil the optimal parameters found for the basis functions with the lower $L$ we perform the additional calculations with the basis set $\chi^i+\gamma_4^s(\zeta)$.
The absolute difference of the results with and without $\gamma_4^s(\zeta)$ is found to be an order of magnitude less than the current uncertainty.
Finally, for the basis set $\chi_0$ it was verified that the GRECP approximation does not affect the optimal parameters for the basis functions.
It turns out that the functions determining the dependence of EA on $\zeta$, evaluated with the $H^{\mathrm{GRECP}}$ and $H^{\mathrm{DC}}$ Hamiltonians for the basis set $\chi_0 + \gamma^s_1(\zeta)$, have extremes at the same values of $\zeta$.
Thus, we extend the uncertainty associated with the GRECP optimized basis set to the calculations based on the other Hamiltonians.
\end{document}